\documentclass[twocolumn,showpacs,preprintnumbers,amsmath,amssymb]{revtex4}

\usepackage[dvips]{graphicx}
\def\gsim{\buildrel {\textstyle >}\over {_\sim}}
\def\lsim{\buildrel {\textstyle <}\over {_\sim}}

\def\Vec#1{\mbox{\boldmath $#1$}}

\begin{document}

\title{Simulation study of earthquakes based on the two-dimensional Burridge-Knopoff model \ 
with the long-range interaction}
\author{Takahiro Mori and Hikaru Kawamura}
\affiliation{Department of Earth and Space Science, Faculty of Science,
Osaka University, Toyonaka 560-0043,
Japan}
\date{\today}
\begin{abstract}

 Spatiotemporal correlations of the two-dimensional spring-block
 (Burridge-Knopoff) models of earthquakes with the long-range inter-block interactions are  extensively studied by means of numerical computer simulations. 
The long-range interaction derived from an elastic theory, which takes account of the effect of the elastic body adjacent to the fault plane, falls off with distance $r$ as $1/r^3$. Comparison is made with the properties of the corresponding short-range models studied earlier. Seismic spatiotemporal correlations of the long-range models generally tend to be weaker than those of the short-range models. The magnitude distribution exhibits a ``near-critical'' behavior, {\it i.e.\/}, a power-law-like behavior close to the Gutenberg-Richter law, for a wide parameter range with its $B$-value, $B\simeq 0.55$, insensitive to the model parameters, in sharp contrast to that of the 2D short-range model and those of the 1D short-range and long-range models where such a ``near-critical'' behavior is realized only by fine-tuning the model parameters. In contrast to the short-range case, the mean stress-drop at a  seismic event of the long-range model is nearly independent of its magnitude, consistently with the observation. Large events often accompany foreshocks together with a doughnut-like quiescence as their precursors, while they hardly accompany aftershocks with almost negligible seismic correlations observed after the mainshock.

\end{abstract}


\maketitle

\section{INTRODUCTION}

An earthquake is a stick-slip dynamical instability of a pre-existing
fault driven by the motion of a tectonic plate \cite{Scholz02,Scholz98}. While an earthquake is a complex phenomenon, certain empirical laws such 
as the Gutenberg-Richter (GR) law and the Omori law concerning its statistical properties are known to hold.
Understanding the origin of such statistical properties of earthquakes is one of important issues left in earthquake studies. As a useful tool in such studies, many researchers have used the so-called spring-block model originally proposed by Burridge and Knopoff (BK) \cite{Burridge}.
In this model, an earthquake fault is simulated by an assembly of blocks,
 each of which is connected via the elastic springs to the neighboring
 blocks and to the moving plate. All blocks are subject to the friction
 force, the source of the nonlinearity in the model, which eventually
 realizes an earthquake-like frictional instability. While the
 spring-block model is obviously a crude model to represent a real
 earthquake fault, its simplicity enables one to study its statistical
 properties with high precision.

 Carlson, Langer and others \cite{CL89a,CL89b,Carletal91,Carl91a,Carl91b,Carletal94}  studied the statistical
 properties of the 1D and 2D BK models quite
 extensively, paying particular attention to the magnitude distribution
 of earthquake events. The spring-block model has also been extended in
 several ways, {\it e.g.\/}, taking account of the effect of 
 viscosity \cite{ML93,Shaw94,DA04}, modifying the form of the
 friction force \cite{ML93,Shaw95,DA04}, driving
 the system only at one end of the system \cite{Vieira92}, or by
 incorporating the rate- and state-dependent friction law
 \cite{Ohm07}. The present authors studied in the previous papers 
the statistical  properties of the 1D and 2D BK models, focusing on their {\it spatiotemporal correlations\/} \cite{MK05,MK06,MK08}. These studies have  revealed several interesting features of the 1D and 2D BK models. 

 Meanwhile, the BK models studied in most of the previous works assumed that the inter-block interaction works only between  nearest-neighboring blocks. This corresponds to the situation where a thin isolated plate is subject to the friction force and is driven by shear force  \cite{CC06}. However, a real fault is not necessarily a thin isolated plate, and the elastic body extends in a direction away from the fault plane. Considering the effect of such an extended elastic body adjacent to the fault plane amounts to considering the effective inter-block interaction to be {\it long-ranged\/}. In order to make the model more realistic, it is important to take account of effect of the long-range interaction, together with the effect of the dimensionality of the fault. In this connection, we note that, in the study of thermodynamic phase transition in equilibrium, it has been wellknown that the spatial dimensionality and the range of the interaction are major elements affecting the universality class of the transition. 

 Hence, in the present paper, we study the statistical properties of the 2D BK model {\it with the long-range inter-block interaction\/} derived from an elastic theory, in comparison with those of the BK models with the short-range (nearest-neighbor) interactions studied earlier, in order to get information how the long-range nature of the interaction, expected to arise from the elastic properties of the crust adjacent to the fault plane, affects the statistical properties of earthquakes.  

 We assume that the 3D elastic body, where the 2D BK models with the long-range interaction is supposed to lie, are isotropic, homogeneous and infinite. A fault surface is assumed to be a plane lying in this elastic body and to slip along one direction only. As a further simplification, 
we adopt a static approximation for an elastic equation of motion describing the elastic body. This assumption is justified when the velocity of the seismic-wave propagation is high enough compared with the velocity of the seismic-rupture propagation. As shown in the appendix, these assumptions give rise to a spring constant between blocks decaying with their distance $r$ as $1/r^3$.

 Certain properties of the BK model with the long-range interaction, or the
 BK model extended in the direction orthogonal to the fault plane, were
 already studied. These include the 2D BK model extended in the
 direction orthogonal to the fault plane \cite{Myersetal96},
 the 2D cellular automaton version of the BK model with the long-range
 interaction decaying as $1/r^3$ \cite{Runetal95}. In particular, Xia {\it et al\/} recently studied the 1D BK model with a variable range interaction where a block is  connected to its $R$ neighbors with a rescaled spring constant
 proportional to $1/R$ \cite{Xiaetal05,Xiaetal07}. The type of the long-range model considered by Xia {\it et al\/} may be regarded as a mean-field type, since the model reduces to the mean-field infinite-range model in the $R\rightarrow \infty$ limit. 

 In the present  paper, we extend our previous studies on the spatiotemporal correlation properties of the short-range BK models \cite{MK05,MK06,MK08}, we investigate the spatiotemporal  correlation properties of the 2D BK model with the long-range power-law interaction {\it derived from an elastic theory\/} which is expected to capture the effect of the elastic body adjacent to the fault plane. Our work can also  be regarded as an extension of the recent work of Xia {\it et al\/} \cite{Xiaetal05,Xiaetal07}: First, we extend the model dimensionality from 1D to more realistic 2D. Second, we consider the long-range interaction derived from an elastic theory, decaying as a power law with distance, which is different from the mean-field-type long-range interaction considered in Ref.\cite{Xiaetal05,Xiaetal07}. Third, we calculate various Spatiotemporal correlation functions to further examine the properties of seismicity under the influence of the long-range interaction. In view of the situation that many of the previous works on the BK model were performed for the 1D model, however, we also perform  for comparison a similar numerical analysis complementally for the 1D BK model with the long-range power-law interaction. 

 The present paper is organized as follows. In \S II, we introduce the model
and explain some of the details of our numerical simulation. The results of our simulations on the 2D BK model with the long-range interactions are presented in \S III. We show the results of the event-size distribution, the mean displacement, the mean number of failed blocks and the mean stress drop at a seismic event, together with various types of spatiotemporal correlation functions of seismic events, including the local recurrence-time distribution, the seismic time-correlation function before and after the mainshock, the time development of the seismic space-correlation function before and after the mainshock, and the time development of the magnitude distribution function before the mainshock. The derivation of the long-range inter-block interaction from an elastic theory is given in Appendix A. The results of our calculation on the 1D BK model with the long-range power-law interaction is also presented in the Appendix B. Finally, \S IV is devoted to summary and discussion.

\section{THE MODEL AND THE METHOD}

 First, we describe the 2D BK model with the nearest-neighbor interaction. 
 The 2D BK model represents a
``fault plane'' by an assembly of blocks, which is taken to be an $x-z$ plane consisting of a 2D square array of blocks containing $N_x$ blocks in the $x$-direction and $N_z$ blocks in the $z$-direction. All Blocks are assumed to move only in the $x$-direction along strike, and 
are subject to the friction force $\Phi$.
Each block is connected with its four nearest-neighbor blocks
 via the springs of the elastic constant $k_c$,
 and is also connected to the moving plate via the spring
 of the elastic constant $k_p$.

 In the simplest case where the interaction works only between the nearest-neighbor blocks in a spatially isotropic manner, the equation of motion of the block at site $(i,j)$ is given by 
\begin{equation}
\begin{array}{ll}
m \ddot U_{i,j}=k_p (\nu ' t'-U_{i,j}) + k_c (U_{i+1,j}+U_{i,j+1} \ \ \
 \ \
\\ \ \ \ \ \ \ +U_{i-1,j}+U_{i,j-1}-4U_{i,j})-\Phi (\dot U_{i,j}),
\end{array}
\end{equation}
where $m$ is the mass of a block, $t'$ is the time, $U_{i,j}$ is the
displacement  along the $x$-direction of the  block at site $(i,j)$, and $\nu '$ is the loading rate representing the speed of the plate.
The equation is made dimensionless in the same way as in \cite{MK06}, {\it i.e.\/}, the time $t'$ is measured in units of the characteristic 
frequency $\omega =\sqrt{k_p/m}$ and the displacement $U_{i,j}$ in units of
the length $L=\Phi(0)/k_p$, $\Phi(0)$ being a static friction. Then,
the equation of motion can be written  in the dimensionless form as
\begin{equation}
\begin{array}{ll}
\ddot u_i=\nu t-u_{i,j}+l^2(u_{i+1,j}+u_{i,j+1} \ \ \ \ \ \\
\ \ \ \ \  +u_{i-1,j}+u_{i,j-1}-4u_{i,j})-\phi (\dot u_i),
\end{array}
\end{equation}
where $t=t'\omega $ is the dimensionless time, 
$u_{i,j}\equiv U_{i,j}/L$ is the dimensionless displacement of the 
block ($i,j$), 
$l \equiv \sqrt{k_c/k_p}$ is the dimensionless stiffness parameter, 
$\nu =\nu '/(L\omega)$ is the dimensionless loading rate, and  
$\phi(\dot u_i) \equiv \Phi(\dot U_i)/\Phi(0)$ is the dimensionless friction 
force.

The nearest-neighbor model mentioned above neglects 
the effect of the elastic body in a direction away from the fault. 
 As shown in  Appendix A, taking account of this effect amounts to taking the inter-block interaction to be long-ranged. The interaction between the two blocks at sites ($i,j$) and ($i^{\prime},j^{\prime}$) is given in the dimensionless form by
\begin{equation}
\left(l^2_x\frac{|i^{\prime}-i|^2}{r^5}+l^2_z\frac{|j^{\prime}-j|^2}
{r^5}\right)(u_{i^{\prime},j^{\prime}}-u_{i,j}), 
\end{equation}
which falls off with distance $r$ as $1/r^3$. Then,
the equation of motion of the 2D long-range can be written as
\begin{equation}
\begin{array}{ll}
\ddot{u}_{i,j}=\nu t-u_{i,j}
\ \ \ \ \\ \ \ \ \
+ \sum_{(i^{\prime},j^{\prime}) \ne (i,j)}
\left(l^2_x\frac{|i^{\prime}-i|^2}{r^5}+l^2_z\frac{|j^{\prime}-j|^2}
{r^5}\right)(u_{i^{\prime},j^{\prime}}-u_{i,j}) - \phi(\dot{u}_{i,j}). 
\end{array}
\end{equation}
If one restricts the range of interaction to nearest neighbors and takes the spatially anisotropic spring constant to be isotropic, $l_x=l_z=l$, one recovers the isotropic nearest-neighbor model described by Eq.(2). 

The ``isotropy'' assumption $l_x=l_z$ is equivalent to putting the Lame's 
constant to vanish, $\lambda =0$. In fact, the possible effect of such 
spatial anisotropy of the 2D BK model  was studied within the
nearest-neighbor interaction in our previous paper \cite{MK08}. It was observed that the property of the anisotropic
model was close to the corresponding isotropic model characterized by
the {\it mean\/} spring constant $l=(l_x+l_z)/2$ so that the spatial
anisotropy did not cause any qualitative new feature on the statistical
properties of the model. Thus, in the present paper, we put $l_x=l_z=l$ for simplicity. The investigation of the recurrence-time distribution  of the anisotropic model with $l_x\neq l_z$ was recently made in Ref.\cite{Hasumi07} within the nearest-neighbor interaction.

 In the present paper, we also discuss in the appendix the properties of the 1D BK model with the long-range interaction, to clarify the role of the model dimensionality and to make comparison with the previous works on the various 1D BK models. We derive the 1D BK model with the long-range interaction from the corresponding 2D model by imposing the constraint that the systems is completely rigid along the $z$-direction corresponding to the depth direction, {\it i.e.\/}, $u(x,z,t)=u(x,t)$. As shown in Appendix A, this yields an effective inter-block interaction decaying with distance $r$ as $1/r^2$,
\begin{equation}
l^2\frac{1}{|i-i^{\prime}|^2}(u_{i^{\prime}}-u_{i}). 
\end{equation}

Then, the equation of motion of the 1D BK model may be given in the dimensionless form by
\begin{equation}
\begin{array}{ll}
\ddot{u}_i=\nu t-u_i+
l^2 \sum_{i^{\prime} \ne i}
\frac{u_{i^{\prime}}-u_{i}}{|i-i^{\prime}|^2}
-\phi(\dot{u}_{i,j}). 
\end{array}
\end{equation}

As the form of the friction force $\phi$, 
we use a simple velocity-weakening friction force which is a single-valued
function of the velocity. As its explicit functional form, we 
use the form introduced by Carlson and Langer \cite{Carletal91},

\begin{equation}
\phi(\dot u) = \left\{ 
             \begin{array}{ll} 
             (-\infty, 1],  & \ \ \ \ {\rm for}\ \  \dot u\leq 0, \\ 
              \frac{1-\sigma}{1+2\alpha \dot u/(1-\sigma )}, &
             \ \ \ \ {\rm for}\ \  \dot u>0, 
             \end{array}
\right.
\end{equation} 
where the friction force immediately drops to $1-\sigma$ on sliding, and
decays toward zero with a rate proportional to the parameter $\alpha$ as
the velocity increases. The back-slip is inhibited by imposing an
infinitely large friction for $\dot u_i<0$, {\it i.e.\/}, 
$\phi(\dot u<0)=-\infty $. 
This friction force represents the {\it velocity-weakening\/} friction force.
Although real friction force is of course more complex, not depending on the velocity alone \cite{Scholz02}, we use the friction force (7) for simplicity.

 The friction force is characterized by the two 
parameters, $\sigma$ and $\alpha$. The former, $\sigma$, 
represents an instantaneous drop of the friction force
at the onset of the slip, while the latter, $\alpha$, 
represents the rate of the friction force getting weaker
on increasing the sliding velocity. The $\alpha = 0$ case represents the
simplest Coulomb friction law  where the friction force instantaneously
drops from the static value $1$ to its dynamical value $1-\sigma$  as
soon as the block begins to slide, and is kept constant on sliding
irrespective of the velocity. The $\alpha =\infty$ case also corresponds
to the another Coulomb friction law where the dynamical friction
immediately drops to zero on sliding.   In addition to these frictional
parameters, the model possesses one more material parameter, an elastic
parameter $l$. 

 In the present paper, we try to cover a rather wide range of the parameter
$\alpha$ in the range $\alpha=[0, \infty]$, and systematically
examine the $\alpha$-dependence of the results.

 We also assume the loading rate $\nu$ to be infinitesimally small, and put 
$\nu=0$ during an earthquake event, a very good approximation 
for real faults \cite{Carletal91}. Taking this limit
ensures that the interval time during successive earthquake events can
be measured in units of $\nu^{-1}$ irrespective of particular values of
$\nu$. 

 A seismic event begins when the accumulated stress exceeds a static friction at one of the blocks in the system. Due to the effect of nonzero $\sigma$, the block begins to move with a finite acceleration, which may (or may not) propagate to the neighboring blocks. The succession of such propagating motion of blocks is regarded as a seismic event. The event is terminated when all blocks in the system come to rest again. The displacement of each block at an event is measured by the displacement of that block during the beginning and the end of this event. The condition of an infinitesimal $\mu$ guarantees that no other event is triggered elsewhere in the system during the ongoing event.

 Numerical details are the same as in \cite{MK06}.
We solve the equation of motion (4) or (6)  
by using the Runge-Kutta method of the fourth 
order, the width of the time discretization $\Delta t$ being
$\Delta t =10^{-3}$ in most cases. The long-range interaction is summed over all blocks contained in the system. 
Total number of $10^5 \sim 10^7$  events are generated in each run, which are used to perform various averagings. The initial position of each block $u_i(0)$ is generated randomly according to the uniform distribution in the interval [0,0.02], with the zero initial velocity $\dot u_i(0)=0$.
In calculating the observables, initial $10^5$ events are discarded as transients. We judge whether the system reaches a stationary state by monitoring the stability of the magnitude distribution function (to be defined in detail below).

  In the 2D BK model, we follow \cite{Carl91b} and impose periodic  boundary condition in the $x$-direction and free boundary condition in the $z$-direction, regarding the  $z$-direction as the depth  direction. For the most part of our calculation, the system size is taken to be $N_x=160$ and $N_z=80$ (or $N_x=60$ and $N_z=60$).  In the 1D BK model studied in Appendix B, we impose periodic boundary condition.

\section{THE SIMULATION RESULTS}

In this section, we show the results of our numerical
simulations on the 2D BK model with the long-range interaction for various observables.

\subsection{THE MAGNITUDE DISTRIBUTION}

We define the magnitude of an event of the 2D BK model, $\mu$, as a
logarithm of its moment $M$,
\begin{equation}
\mu= \ln M = \ln \left( \sum_{i,j} \Delta u_{i,j} \right),
\end{equation}
where $\Delta u_{i,j}$ is the total displacement during an event of the  block at site ({$i,j$})  and the sum is taken over all blocks involved in
the event.

\begin{figure}[ht]
\begin{center}
\includegraphics[scale=0.6]{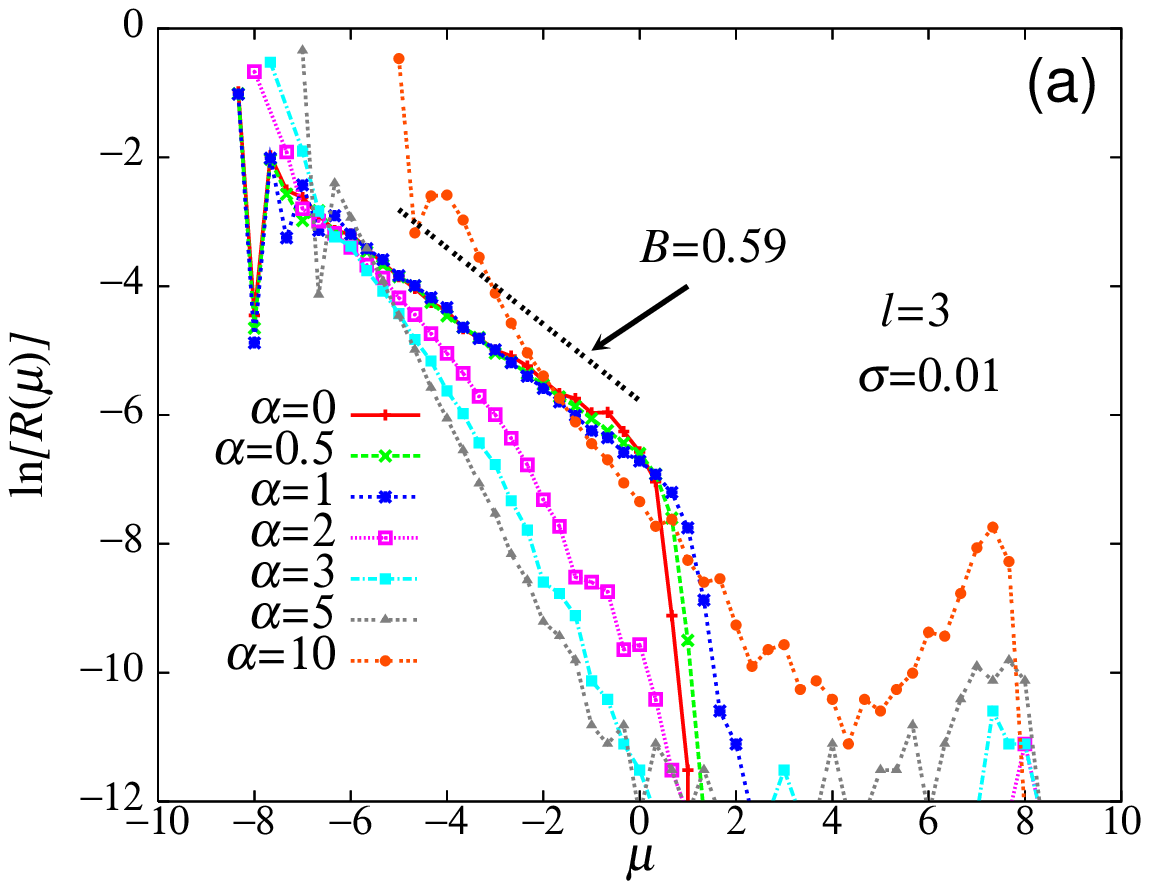}
\includegraphics[scale=0.6]{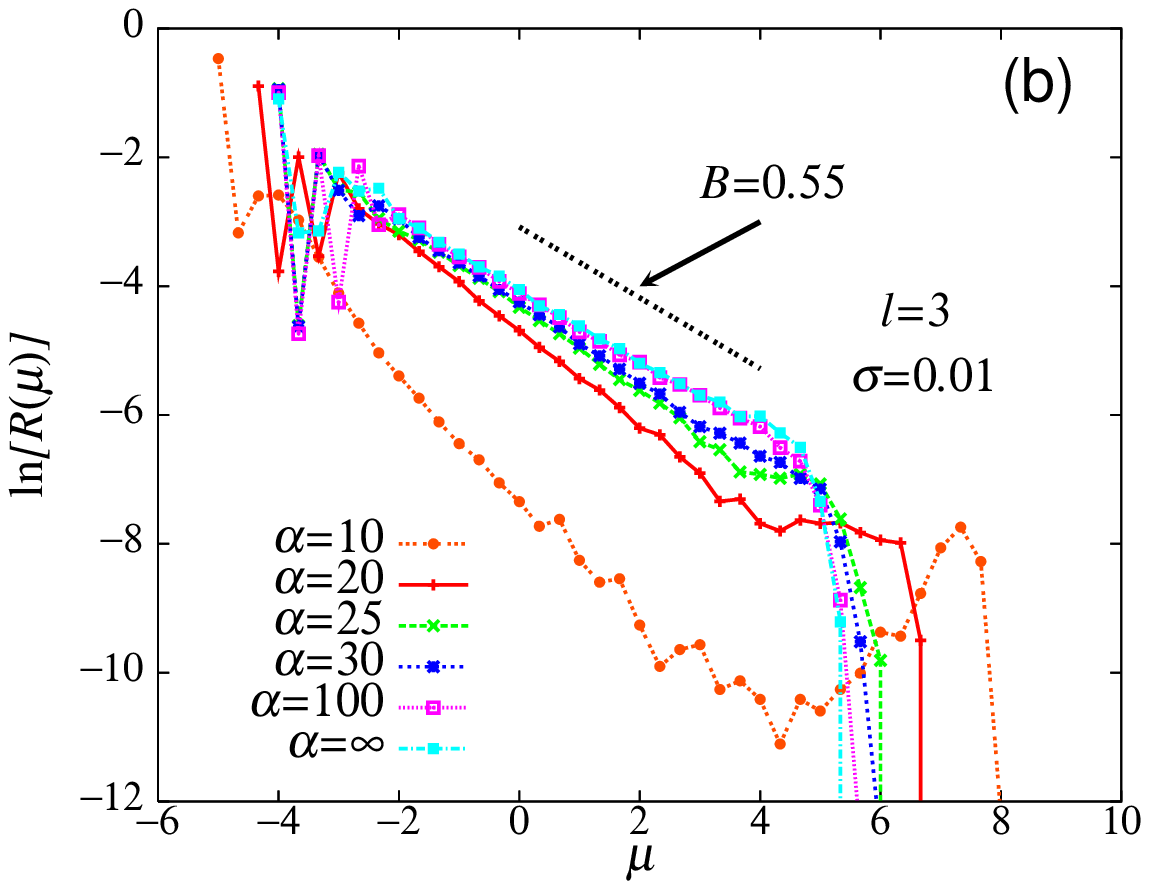}
\end{center}
\caption{
The magnitude distribution $R(\mu)$ of earthquake events of the 2D long-range
 BK model for the  parameters $l=3$ and $\sigma=0.01$. Fig.(a) represents $R(\mu)$ for  smaller values of the frictional parameter $0\leq \alpha \leq 10$, while Fig.(b) represents $R(\mu)$ for larger values of the frictional parameter $10\leq \alpha \leq \infty$. The system size is $60\times 60$.
}
\end{figure}

 Figs.1(a) and (b) exhibit the computed magnitude distribution function
 $R(\mu)$ for smaller and larger values of $\alpha$, {\it i.e.\/}, (a) $0\leq \alpha \leq 10$ and (b) $10\leq \alpha \leq \infty$. The magnitude  distribution $R(\mu){\rm d}\mu$ represents the rate of events with  their magnitudes in the range [$\mu, \mu + {\rm d}\mu$]. 
 In the range $\alpha \lsim 1$, only small events
 of $\mu \lsim 2$ occur. As can be seen from Fig.1(a), $R(\mu)$ for smaller $\alpha \lsim 1$ exhibits  a  near straight-line ``near-critical'' behavior over a certain magnitude range, and drops off sharply at larger magnitudes. The associated $B$-value  is estimated in the range $0\lsim \alpha \lsim 1$ to be $B\simeq 0.59$ from the slope of this straight line, which is rather insensitive to the change of the $\alpha$-value.  Of course, the observed  behavior cannot be regarded as truly critical, since $R(\mu)$ drops off sharply beyond the threshold magnitude.

\begin{figure}[ht]
\begin{center}
\includegraphics[scale=0.65]{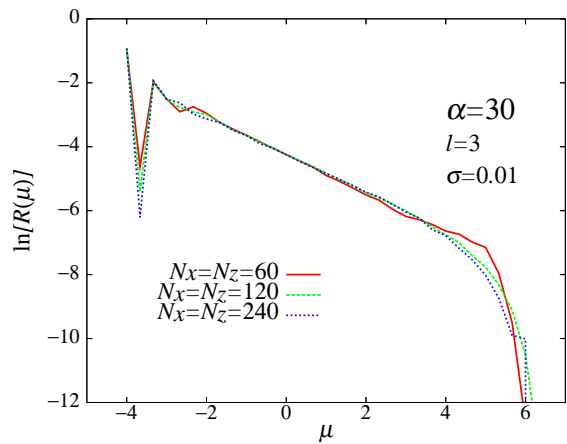}
\end{center}
\caption{
The system-size dependence of the magnitude distribution $R(\mu)$ of earthquake
 events of the 2D long-range BK model
 for the  parameters $\alpha=30$, $l=3$ and $\sigma=0.01$. A sharp fall-off observed at larger magnitudes $\mu\gsim 5$ is not a finite-size effect.
}
\end{figure}

  At $\alpha \gsim 2$, large earthquakes of their magnitudes $\mu \simeq 8$ suddenly appear, while  earthquakes of intermediate magnitudes, say, $2\lsim \mu \lsim 6$,  remain rather scarce. It means that large and small earthquakes are
 well separated at $\alpha\simeq 2$. Such a sudden appearance of large
 earthquakes at $\alpha =\alpha_{c1}\simeq 2$ coexisting with smaller
 ones has a feature of ``discontinuous transition''. This feature is common to the case of the corresponding 2D short-range model \cite{MK08}.  On increasing $\alpha$ further, earthquakes of intermediate
 magnitudes gradually increase their frequency.
In the range of $2\lsim \alpha \lsim 20$,
 $R(\mu)$ exhibits a ``supercritical'' behavior, {\it i.e.\/}, exhibits a pronounced peak structure at a larger magnitude  deviating from the GR law, while it
 still exhibits a near straight-line behavior corresponding to the GR law at
 smaller magnitudes. The existence of a distinct peak structure at a larger magnitude  suggests that large earthquakes are more or
 less characteristic. Such a behavior of  $R(\mu)$ is sometimes called ``supercritical'', since $R(\mu)$ bends up at larger magnitudes (though it eventually falls off at still larger magnitudes). 

\begin{figure}[ht]
\begin{center}
\includegraphics[scale=0.58]{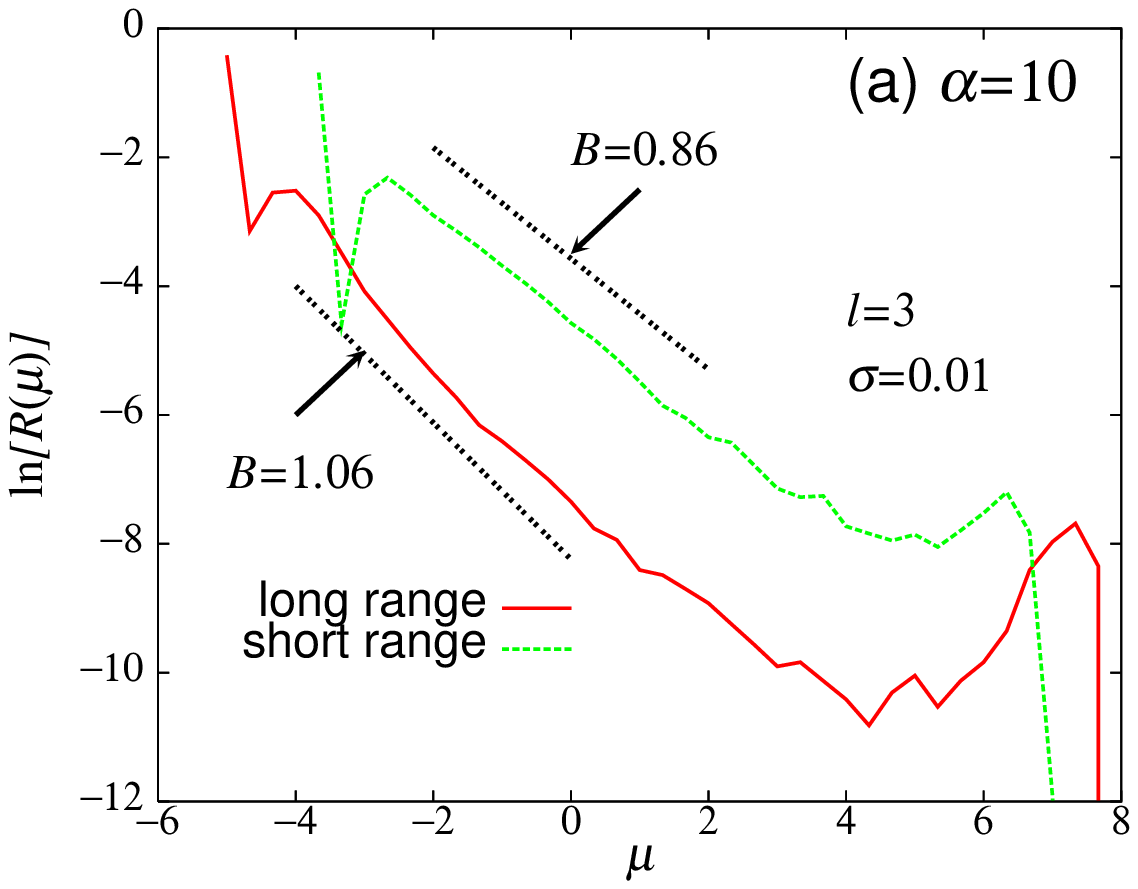}
\includegraphics[scale=0.58]{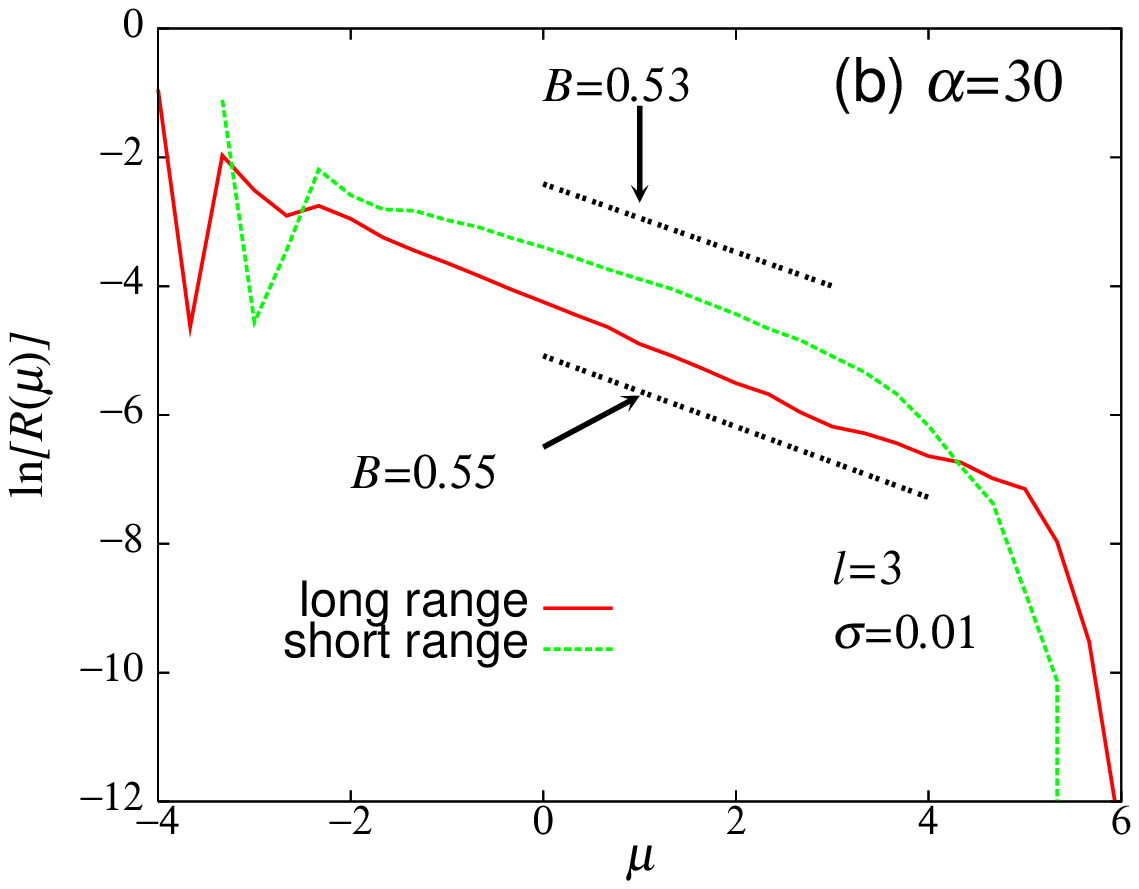}
\end{center}
\caption{
Comparison of the magnitude distribution $R(\mu)$ of earthquake events of the 2D BK models with 
 the long-range and the short-range interactions. Fig.(a) represents the case of $\alpha =10$, 
 while Fig.(b) the case of $\alpha =30$, with $l=3$ and $\sigma=0.01$ being fixed. 
}
\end{figure}

  As $\alpha$ increases further, a characteristic peak  becomes less pronounced and eventually vanishes  at around $\alpha \simeq 25$.
 $R(\mu)$ exhibits again a near straight-line near-critical behavior over a wide magnitude range: See Fig.1(b). At $\alpha =\alpha_{c2}\simeq 25$, the associated $B$-value estimated from the slope of this straight line is $B\simeq 0.55$. The  change from the supercritical to the near-critical behaviors at  $\alpha=\alpha_{c2}\simeq 25$ is continuous, in  contrast to the discontinuous one observed at  $\alpha=\alpha_{c1}\simeq 2$. 
 A very interesting observation here is that {\it such a straight-line near-critical behavior persists even if $\alpha$ is further increased up to $\alpha=\infty$, and that the associated $B$-value is robust against the change of $\alpha$\/}. It should be noticed that this straight-line behavior of $R(\mu)$ cannot be regarded as a truly critical one, since $R(\mu)$ drops off sharply at very large magnitudes. This sharp fall-off of $R(\mu)$ observed at larger magnitudes $\mu\gsim 5$ is not a finite-size effect, as can clearly be seen from Fig.2. 

 Such a near-critical behavior realized over a wide parameter range $\alpha\gsim 25$ is in sharp contrast to the behavior of the corresponding short-range model where $R(\mu)$ at larger $\alpha > \alpha_{c2}$ exhibits  a down-bending ``subcritical'' behavior, while a straight-line near-critical behavior is realized only by fine-tuning the $\alpha$-value to a special value $\alpha\simeq \alpha _{c2}$. The robustness of the near-critical behavior of $R(\mu)$ observed in the 2D long-range model might have an important relevance to real seismicity.

\begin{figure}[ht]
\begin{center}
\includegraphics[scale=0.58]{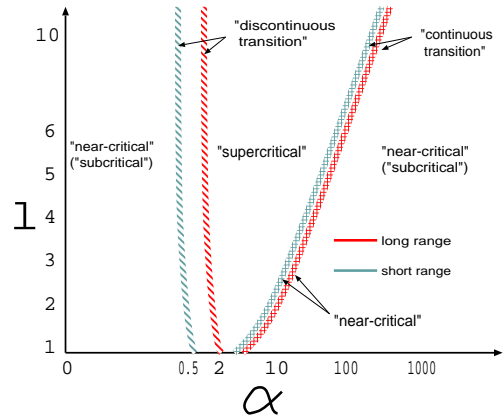}
\end{center}
\caption{
The phase diagram of the 2D BK models with the long-range and the short-range interactions in the frictional-parameter $\alpha$ versus elastic-parameter $l$ plane.  The parameter $\sigma$ is set to $\sigma=0.01$. To draw a phase diagram, the parameter range $0\leq \alpha \leq \infty$ and $1\leq l \leq 10$ is studied by simulations.
}
\end{figure}

 In order to further illustrate the difference between the long-range and the  short-range models, we compare in Fig.3 $R(\mu)$ of the long-range and 
 the short-range models. Fig.3(a) represents the case of $\alpha=10$ in the supercritical regime  $\alpha_{c1} <\alpha < \alpha_{c2}$, while Fig.3(b) represents the case of $\alpha=30$ in the near-critical regime $\alpha > \alpha_{c2}$. As can be seen from the figures, $R(\mu)$ of the long-range model exhibits much more pronounced straight-line behavior mimicking the GR-law over a wider magnitude range, as compared with $R(\mu)$ of the short-range model.

 In Fig.4, we summarize the behavior of $R(\mu)$ in the form of a
 ``phase diagram'' in the frictional-parameter $\alpha$ versus the
 elastic-parameter $l$ plane for the case of $\sigma=0.01$. As can be seen from the figure, the phase
 diagram of the long-range model consists of three distinct regimes,  two of which are near-critical regimes and one is a supercritical regime. The
 ``phase boundary'' between the smaller-$\alpha$  near-critical regime
 and the supercritical regime represents a ``discontinuous transition'', while the one between the larger-$\alpha$
 near-critical regime and the supercritical regime represents a
 ``continuous  transition''. The ``transition'' between these
 different ``phases'', {\it i.e.\/}, a near-critical phase for small
 $\alpha$, a supercritical phase for intermediate $\alpha$, and
 another near-critical phase for large $\alpha$, is primarily dictated
 by the $\alpha$-value. Since the phase boundary in Fig.4
 has a finite slope in the $\alpha$-$l$ plane,  one can also induce the
 near-critical to supercritical transition by increasing the
 $l$-value for a fixed $\alpha$. 

 For comparison, we also show in Fig.4 the corresponding phase boundary of the short-range model reported in Ref.\cite{MK08}. As can be seen from the figure, the phase diagram of both the long-range and the short-range models are qualitatively similar. The near-critical phases in the long-range model are replaced by the subcritical phases in the short-range model, and the phase boundaries of the long-range model tend to shift to larger values of $\alpha$ and to smaller values of $l$.

\subsection{THE MEAN DISPLACEMENT, THE MEAN NUMBER OF FAILED-BLOCKS AND THE MEAN STRESS-DROP}

  The size of an earthquake event is usually measured by its magnitude. Other possible measures of event size might be the mean displacement $\Delta \bar u$, the mean number of failed-blocks $\bar N_b$ (corresponding to the size of rupture zone), and the mean  stress-drop $\Delta \bar{\tau}$. In Figs.5(a) and (b), we show the magnitude dependence of the mean displacement and of the mean number of failed-blocks for various values of $\alpha$.  An interesting observation is that the data in the near-critical regimes are grouped into two distinct branches, each corresponding to the small-$\alpha$ and large-$\alpha$ near-critical regions of Fig.4.

 As can be seen from Fig.5(a), the data in the small-$\alpha$ near-critical regime ($\alpha < \alpha_{c1} \simeq 2$) lacks events of larger magnitudes and are characterized by smaller displacement, while those in the large-$\alpha$ near-critical regime ($\alpha > \alpha_{c2} \simeq 25$) are characterized by much larger displacement. All the data of the mean displacement $\Delta \bar u$ in the near-critical regimes collapse, at least approximately,  onto these two curves, which are both linear in the magnitude with a common slope $\simeq 0.01$. Note that this slope is very small, indicating that the mean stress-drop in the near-critical regime hardly depends on the event magnitude. This slope is also an order of magnitude smaller than the corresponding slope observed in the 2D short-range BK model, which was estimated to be about 0.1 \cite{MK08}.

 By contrast, the data in the supercritical regime ($\alpha_{c1} < \alpha < \alpha_{c2}$) exhibit a significantly different behavior. At smaller magnitudes $\mu\lsim 5$, they exhibit a crossover behavior depending on its $\alpha$-value between the two universal near-critical curves: For smaller $\alpha$ close to $\alpha_{c1}$, the data tend to lie closer to the small-$\alpha$ near-critical curve, while for larger $\alpha$ close to $\alpha_{c2}$, the data tend to lie closer to the large-$\alpha$ near-critical curve. The data in the supercritical regime suffer from significant finite-size effects at larger magnitudes $\mu\gsim 5$. The system-size $N$ dependence of the data in the near-critical regime $\alpha=30$ is shown in the inset of Fig.5(a). As can be seen from the inset of Fig.5(a), the data at  larger magnitudes tend to level off as the system-size $N$ is increased.

 The existence of the two near-critical curves and the crossover behavior between them are also clearly visible in Figs.5(b) for the magnitude dependence of the mean number of failed-blocks $\bar N_b$. The two near-critical curves are again both strikingly linear with a common slope $\simeq 0.99$. The system-size $N$ dependence  is shown in the inset of Fig.5(b) for the case of $\alpha=30$. As the system size is increased, the data at larger magnitudes tend to lie closer to a straight line of a slope $\simeq 0.99$.

\begin{figure}[ht]
\begin{center}
\includegraphics[scale=0.54]{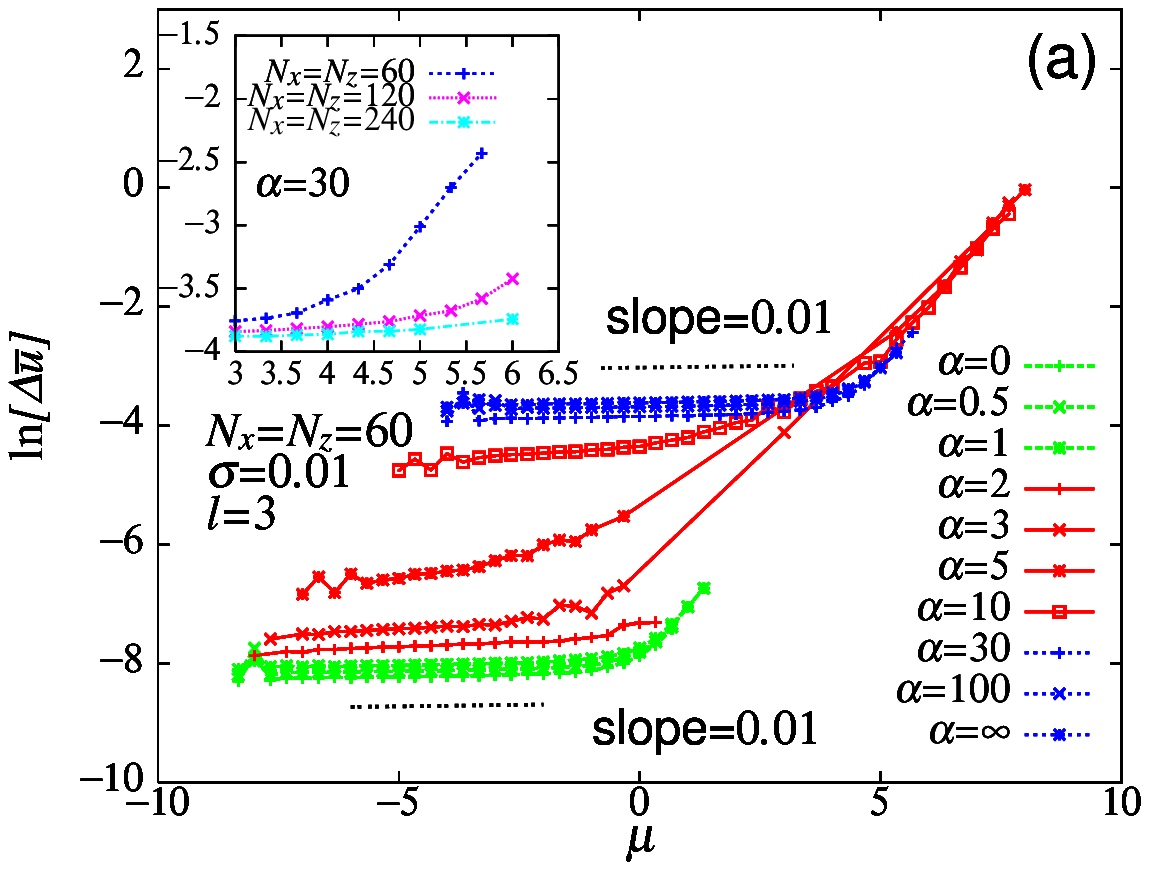}
\includegraphics[scale=0.54]{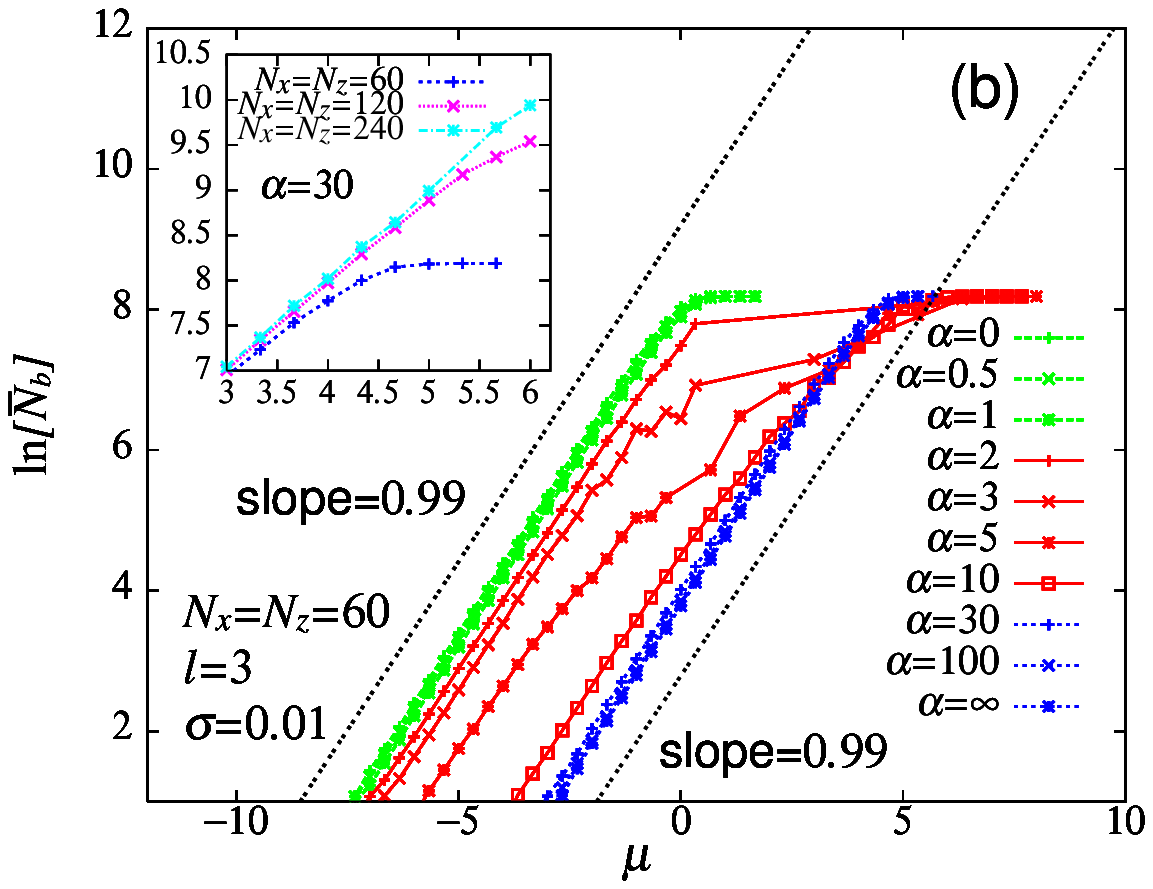}
\includegraphics[scale=0.54]{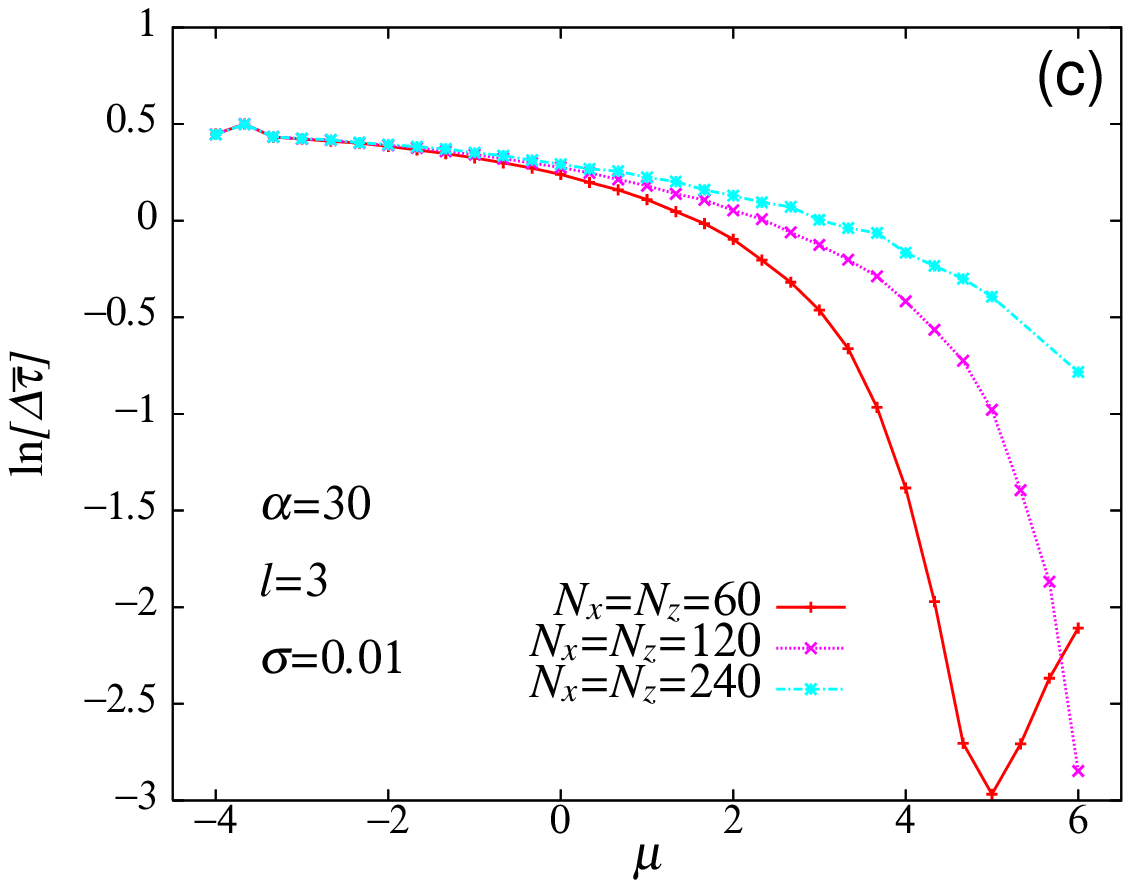}
\end{center}
\caption{
  The magnitude dependence of the mean displacement (a),
 the mean number of failed-blocks (b), and the mean stress-drop (c) of each seismic event of the 2D long-range BK model. In the main panels of Figs.(a) and (b), the frictional-parameter $\alpha$ is varied with fixing the system-size $60\times 60$, while in the insets the system-size $N$ is varied for the case of $\alpha=30$. In Fig.(c), the system-size dependence of the mean stress-drop is shown for the case of $\alpha=30$.   The parameters $l$ and  $\sigma$ are fixed to $l=3$ and $\sigma=0.01$.
}
\end{figure}

In Fig.5(c), we show the magnitude dependence of the mean stress-drop for the case of $\alpha =30$, with varying the system size $N$. Note that, although in the nearest-neighbor BK model the mean stress-drop of an event is essentially identical with (proportional to) the mean displacement of an event \cite{MK08}, such a simple relation between the mean displacement and the mean stress-drop does not hold in the present long-range BK model. Although a significant finite-size effect is observed, there clearly exists a tendency that the magnitude dependence becomes less and less for larger systems. In real seismicity, the mean stress-drop is known to hardly
depend on the event magnitude \cite{Scholz02}. This suggests that the long-range nature of the elastic interaction of the crust might play a role in realizing the near-independence of the stress-drop on the event magnitude. We note that a similar independence was also observed in the mean-field-type 1D long-range BK model studied by Xia {\it et al\/} \cite{Xiaetal07}, and also in the 1D long-range BK model with a power-law interaction studied in Appendix B.

\subsection{THE LOCAL RECURRENCE-TIME DISTRIBUTION}

\begin{figure}[ht]
\begin{center}
\includegraphics[scale=0.65]{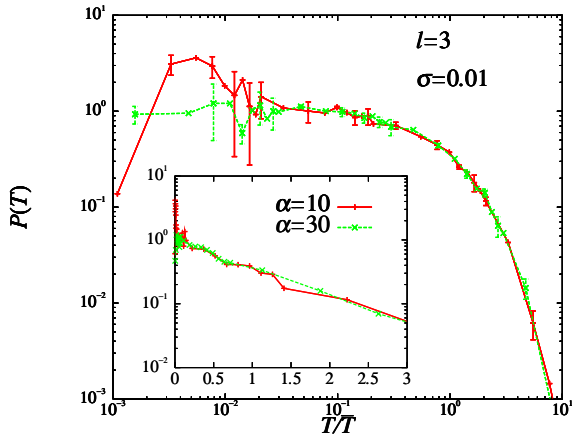}
\end{center}
\caption{
The log-log plot of the local recurrence-time distribution  $P(T)$ of large events of $\mu\geq \mu_c=5$  of the 2D long-range BK
 model for the frictional-parameter $\alpha=10$ and 30, each corresponding to the  ``supercritical'' and ``near-critical'' regimes.
 The parameters $l$ and
 $\sigma$ are fixed  to $l=3$ and $\sigma =0.01$. The system size is
 $160\times 80$. The recurrence time $T$ is normalized by its mean 
 $\bar T$, which is $\bar T=31.5$ and 9.98 for $\alpha=10$
 and 30, respectively.
 The insets represent the semi-logarithmic plots including the tail part
 of the distribution. The tail part shows an exponential behavior for
 both cases of $\alpha=10$ and 30.
}
\end{figure}

  In earthquake prediction, one natural quantity to be
 investigated might be  the distribution law of the recurrence time of large earthquakes. Characteristic  earthquake recurrence  would mean the existence of  characteristic time scales in earthquake recurrence, while critical earthquake recurrence would mean the absence  of such characteristic time scales. Here, we study the nature of earthquake recurrence of the 2D long-range BK model via the {\it local\/} recurrence-time  distribution function. 
  
In Figs.6, we show on a log-log plot the  computed local recurrence-time distribution function $P(T)$ for the cases of $\alpha =10$ (a) and $\alpha=30$ (b), with fixing $l=3$ and $\sigma =0.01$. Each case corresponds to the supercritical and the near-critical regimes, respectively. The local recurrence time $T$ is recorded when the next event occurs with its epicenter lying within distance $r=5$ from the epicenter of the previous event.
In the insets, the same data are re-plotted on a
semi-logarithmic scale. The recurrence time is normalized by its mean
$\bar T$, which is $\bar T\nu =31.5$ and 
9.98 for $\alpha=10$ and 30, respectively.
As can be seen from the figure, $P(T)$ exhibits an exponential tail at longer times for both cases of $\alpha=10$ and 30, with and without a peak structure at short times.

\begin{figure}[ht]
\begin{center}
\includegraphics[scale=0.65]{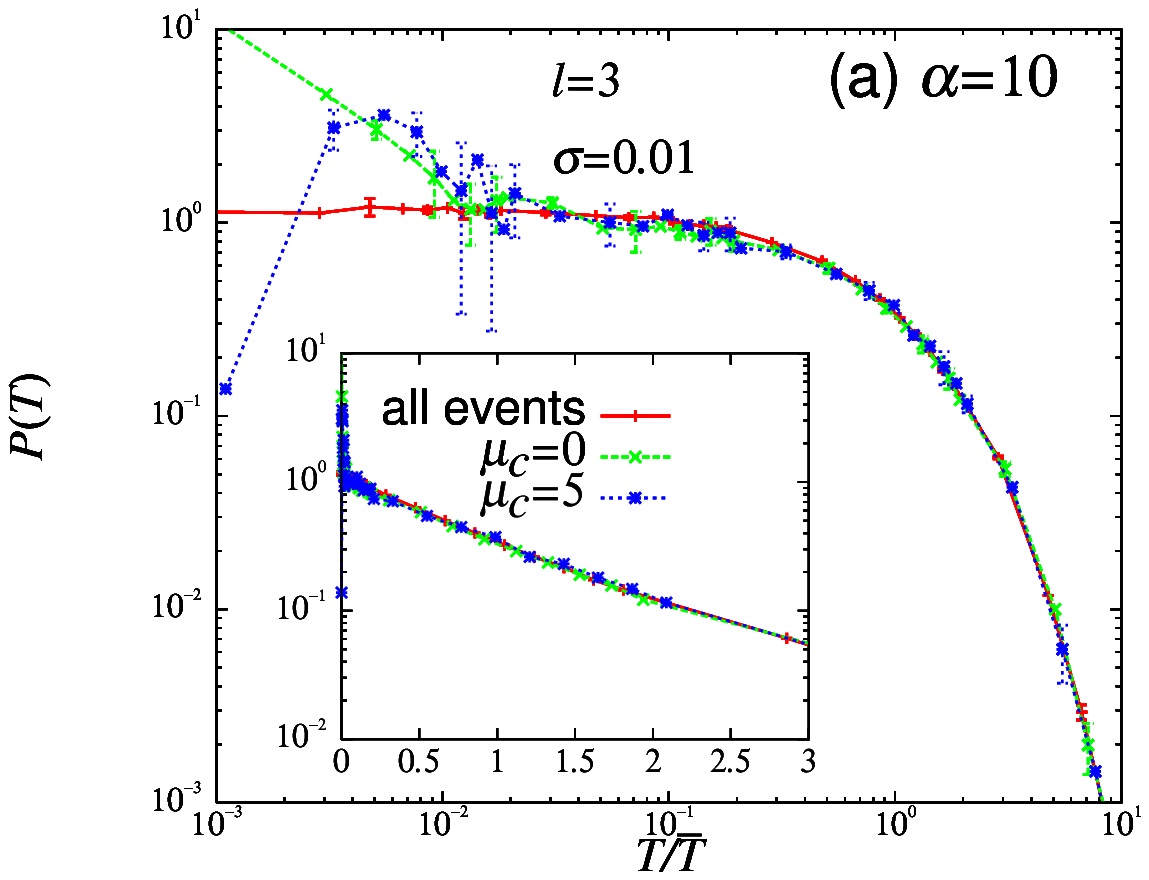}
\includegraphics[scale=0.65]{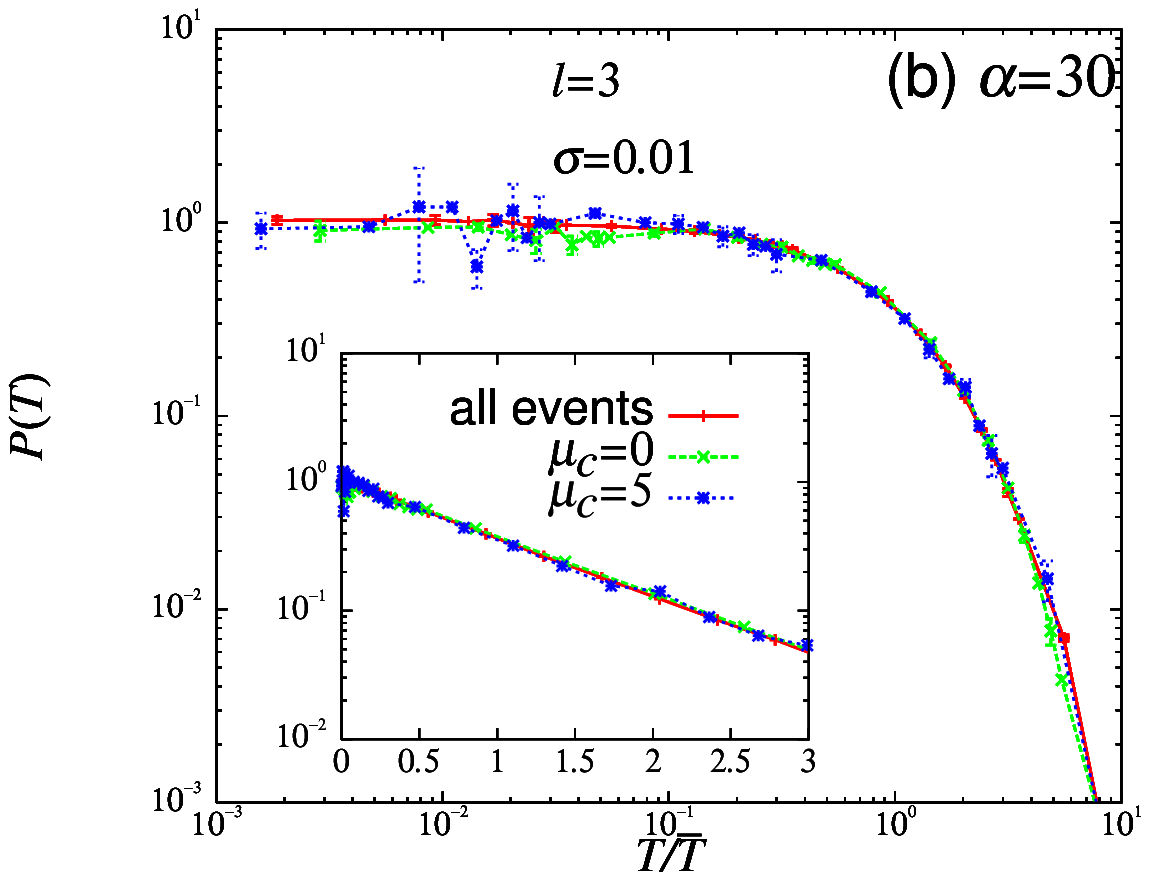}
\end{center}
\caption{
 The log-log plots of the local recurrence-time distribution function $P(T)$ of the 2D long-range BK model, with
 varying the magnitude-threshold $\mu_c$ for the cases of the frictional-parameter  $\alpha=10$ (a) and $\alpha =30$ (b). 
 The insets represent the semi-logarithmic plots including the 
 tail part of the distribution. The mean recurrence time is
 $\bar T\nu =0.0262, 14.3$ and 31.5 (respectively for $\mu_c=-\infty, 0$ and 5)   for $\alpha=10$, and $\bar T=0.0135, 0.37$ and 9.98
 (respectively for $\mu_c=-\infty, 0$ and 3) for $\alpha=30$.
}
\end{figure}

 In Figs.7, we show $P(T)$ for various values of the magnitude-threshold $\mu_c$, including the case of $\mu_c=-\infty$ corresponding to no threshold at all (all events), for the cases of $\alpha=10$ (a) and of $\alpha =30$ (b). Both in the cases of $\alpha=10$ and 30, $P(T)$ robustly exhibits a down-bending behavior  for any choice of $\mu_c$. In the supercritical case of $\alpha=10$, a prominent peak observed at shorter $T$ for $\mu_c=5$ tends to be suppressed as $\mu_c$ is taken smaller. In the near-critical case of $\alpha=30$, no characteristic peak is observed for any choice of $\mu_c$. The appearance of the characteristic peak in $P(T)$ at $T\simeq 0.01\bar T$ in the supercritical regime and for larger events is well correlated with the appearance of the characteristic-peak component in the magnitude distribution $R(\mu)$ of Fig.1. The recurrence-time distribution in real seismicity usually does not exhibit a characteristic peak (see, {\it e.g.\/}, Fig.5 of Ref.\cite{Kawamura06}). Hence, the behavior of $P(T)$ in the near-critical regime seems closer to that of real seismicity.

\begin{figure}[ht]
\begin{center}
\includegraphics[scale=0.65]{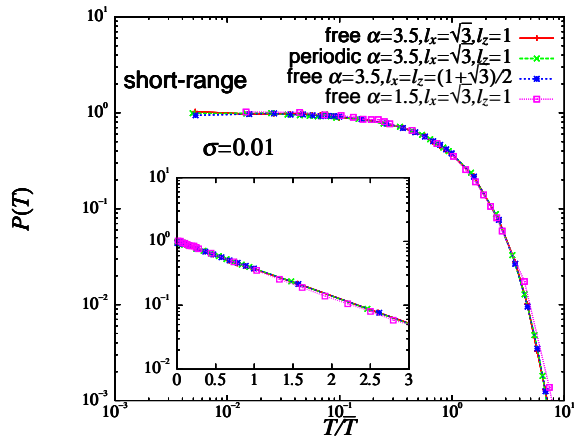}
\end{center}
\caption{
 The log-log plot of the global recurrence-time distribution function  $P(T)$ of the 2D short-range BK model. No constraint is imposed on the magnitude threshold, {\it i.e.\/}, $\mu_c=-\infty$. The system size is $N_x=25$ and $N_z=100$. The cases of $\alpha=1.5$ and 3.5, with $l_x=\sqrt{3}, l_z=1, \sigma=0.01$ and with free boundary conditions applied both in $x$- and $z$-directions, precisely correspond to the cases studied in Ref.\cite{Hasumi07}. 
In these cases, we get $\bar T\nu =0.48$ for $\alpha=3.5$, and $\bar T\nu =1.70$ for $\alpha=1.5$. 
In the case of $\alpha=3.5$, the data taken under periodic boundary conditions applied in both $x$- and $z$-directions as well as the data taken for the isotropic elastic couplings $l_x=l_z=(1+\sqrt{3})/2$ are also given. 
In the former case, we get $\bar T=0.51$, while in the latter case we get $\bar T=0.48$. 
The main panel represents the log-log plot of $P(T)$, while  the inset represents the semi-logarithmic plot including the  tail part of the distribution. In all cases, the computed $P(T)$ exhibits a down-bending behavior, in sharp contrast to Ref.\cite{Hasumi07}. The reason of this deviation is discussed in the text.
}
\end{figure}

 While the present results of $P(T)$ turn out to be qualitatively similar to those of the 2D {\it short-range\/} model calculated by the present authors \cite{MK08}, they differ significantly from the recent result of the recurrence-time (interoccurrence-time) distribution reported by Hasumi for the 2D {\it short-range\/} BK model \cite{Hasumi07}. Hasumi reported that the recurrence-time distribution $P(T)$, defined globally, exhibits either a supercritical, subcritical or critical behavior depending on the $\alpha$-value, which is well correlated with the behavior of the magnitude distribution function $R(\mu)$. Such behaviors of $P(T)$, however, were never observed in our calculation of the 2D BK model either in the short-range nor in the long-range case. In order to trace the cause of this significant deviation from Ref.\cite{Hasumi07}, we further calculated the {\it global\/} recurrence-time distribution  on exactly the same 2D short-range BK model as studied by Hasumi, imposing no constraint on the distance between successive events. The result is shown in Fig.8. First, we closely follow Ref.\cite{Hasumi07} by applying free boundary conditions in both directions on the lattice of size $N_x=25$ and $N_z=100$, assuming the {\it anisotropic\/} elastic parameters $l_x=\sqrt 3$, $l_z=1$, setting the other parameter values to $\sigma=0.01$ and $\alpha=1.5$ or 3.5, and imposing no magnitude-constraint $\mu_c=-\infty$. Precisely under these calculational conditions, Hasumi observed a critical straight-line $P(T)$ for the case of $\alpha=3.5$, and a subcritical up-bending $P(T)$ accompanying a characteristic larger-$T$ peak for the case of $\alpha=1.5$. In sharp contrast to this, we observed here a subcritical down-bending $P(T)$ for either value of $\alpha$: See fig.8.

 In the case of $\alpha=3.5$, we also examined the possible effect of the boundary conditions and of the anisotropy of the elastic constants on $P(T)$ by simulating the model under periodic boundary conditions and with the isotropic elastic constants $l_x=l_z=(l_x+l_z)/2$, to find that the applied boundary conditions and the anisotropy of the elastic constants hardly affect the form of $P(T)$ as shown in Fig.8. 

 In fact, Hasumi included the rise-time of earthquakes in his definition of the  recurrence time, and  assumed an extremely large loading rate, $\nu=10^{-2}$ \cite{Hasumi-pc}. We believe that his choice of unrealistically large loading rate, combined  with his definition of the recurrence time, is the cause of the deviation between our results and those of Ref.\cite{Hasumi07}. As is well known, in real seismicity the loading rate is extremely small, being of order $\nu=10^{-8}-10^{-9}$. Then, with such a  realistic choice of the $\nu$-value, the recurrence-time distribution $P(T)$ of the 2D BK model, either local or global, should behave in the way as reported in the present paper and in Ref.\cite{MK08}, {\it not\/} as reported in Ref.\cite{Hasumi07}, irrespective of whether one includes the rise-time part of earthquakes in the definition of the recurrence time or not.

\subsection{TIME CORRELATIONS OF EVENTS ASSOCIATED WITH THE MAINSHOCK}

\begin{figure}[ht]
\begin{center}
\includegraphics[scale=0.65]{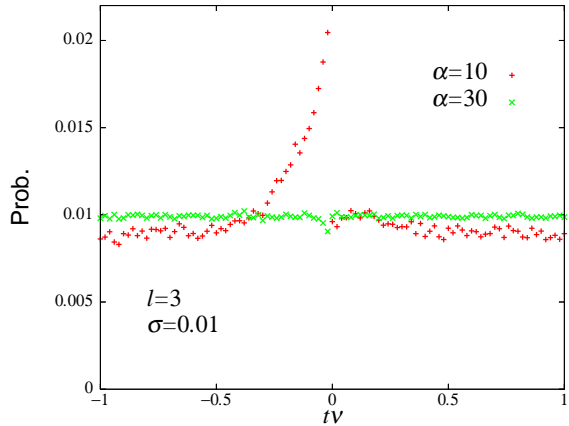}
\end{center}
\caption{
The time correlation function of the 2D long-range BK model
between large events of $\mu_c=5$ (mainshock) occurring at time $t=0$ and
events of arbitrary sizes, dominated in number by small events, occurring at
time $t$ for the cases of $\alpha =10$ and 30. 
The parameters $l$ and $\sigma$ are fixed to $l=3$ and $\sigma =0.01$.
Events of arbitrary sizes occurring within 5 blocks from the epicenter of the
mainshock are counted. 
The negative time $t<0$ represents the time before the mainshock, 
while the positive time $t>0$ represents the time after the mainshock. 
The average is taken over all large events with its magnitude 
$\mu >\mu_c=5$. The system size is  $160\times 80$. 
}
\end{figure}

 In real seismicity, large events often accompany foreshocks and aftershocks. In Fig.9, we show the time correlation function between
large events (mainshock) and events of arbitrary sizes, dominated in
number by small events, for various values of the frictional-parameter
$\alpha$, with fixing $l=3$ and $\sigma=0.01$.
 In the figure, we plot the
mean number of events of arbitrary sizes occurring within 5
blocks from
the epicenter of the mainshock before  ($t<0$) and after ($t>0$) the
mainshock, where the occurrence of the mainshock is taken to be the
origin of the time $t=0$. The average is taken over all large events of
their magnitudes of $\mu \geq \mu_c=5$. The number of events are
counted here with the time bin of $\Delta t\nu =0.02$.

 As can be seen from Fig.9,
 a remarkable acceleration of seismic activity occurs  before the
 mainshock ($t<0$) for $\alpha=10$ corresponding to
 the supercritical regime, while, for $\alpha=30$ corresponding to the near-critical  regime, the time correlation is
 almost absent except for the suppression of seismicity immediately
 before the mainshock. The behavior of the time-correlation function of the 2D long-range model turns out to be similar to those of the corresponding 2D short-range model \cite{MK08}.

\subsection{SPATIAL CORRELATIONS OF EVENTS BEFORE THE MAINSHOCK}

 In this subsection, we examine the time-development of spatial seismic
correlations {\it before\/} the mainshock 
of $\mu >\mu_c =5$. In  Figs.10, we show the spatial seismic correlation functions between  the mainshock and the preceding events of arbitrary size, dominated in number by small events, for several time periods  before the mainshock,
 with fixing $l=3$ and $\sigma=0.01$. Figs.10(a) and (b) represent the cases of $\alpha=10$ in the  supercritical  regime and of $\alpha=30$ in the near-critical  regime, respectively. Insets represent shorter-time behaviors.

\begin{figure}[ht]
\begin{center}
\includegraphics[scale=0.65]{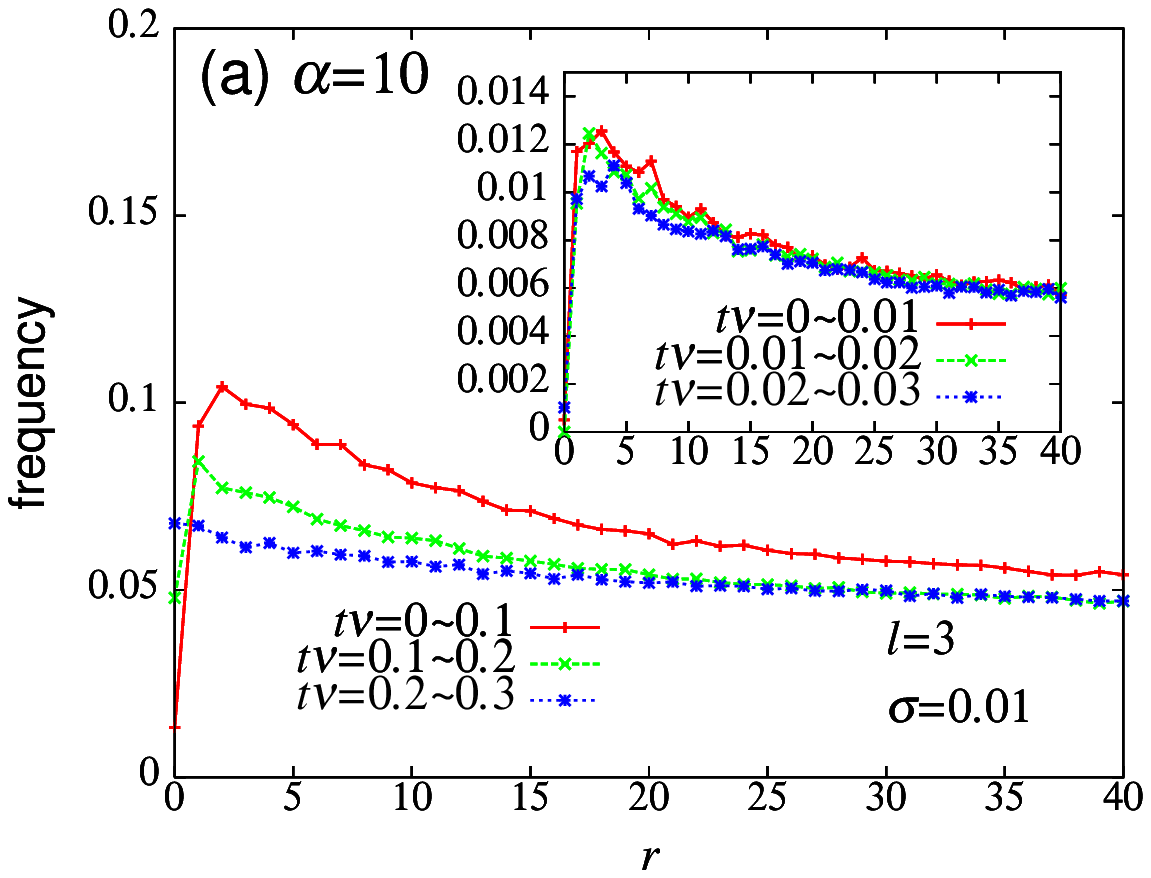}
\includegraphics[scale=0.65]{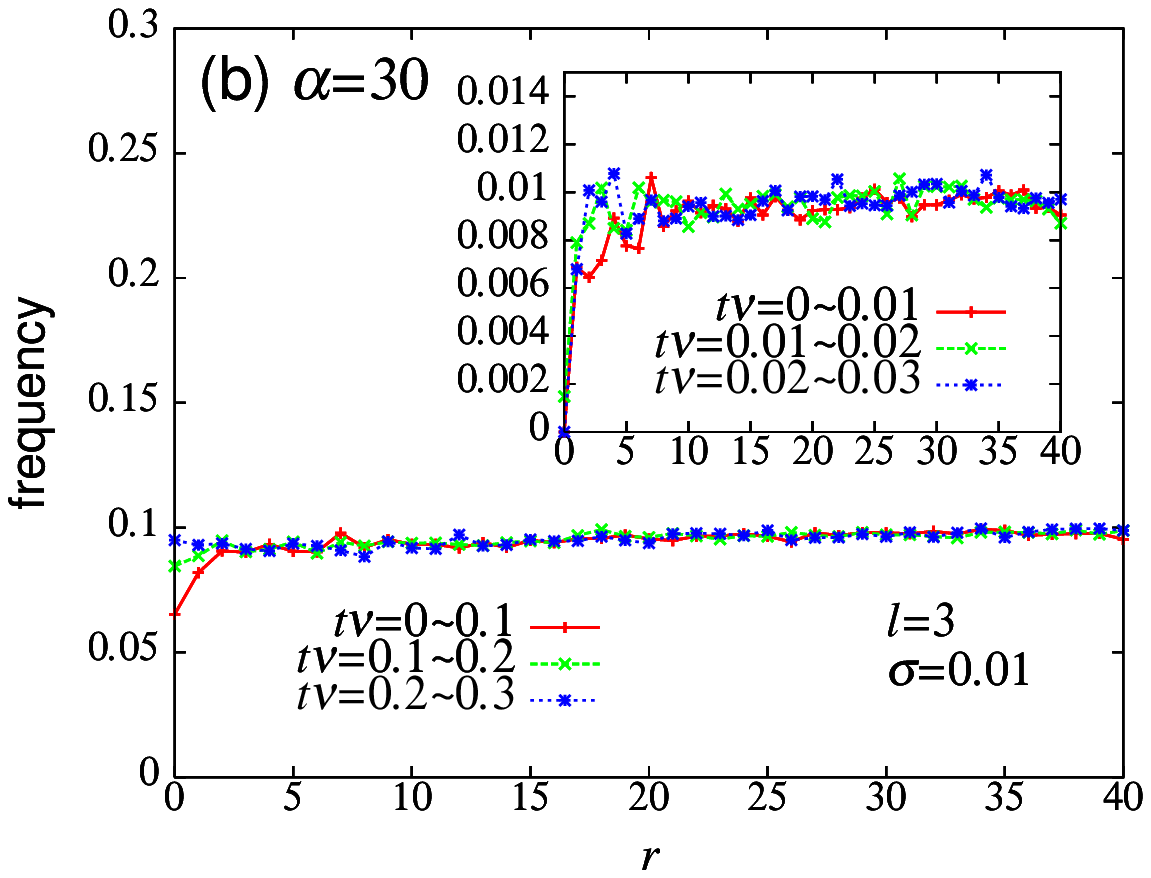}
\end{center}
\caption{
Event frequency before the large event of $\mu >\mu_c=5$
 (mainshock) plotted versus $r$,
the distance from the epicenter of the upcoming mainshock, for several
 time periods before the mainshock in the 2D long-range BK model.
 The frictional-parameter $\alpha$ is 
$\alpha=10$ (a) and $\alpha=30$ (b) with $l=3$ and $\sigma=0.01$, each
 corresponding to the
 ``supercritical'' and ``near-critical'' regimes. The
 system size is $160\times 80$. The insets represent similar plots at
 shorter times.
}
\end{figure}

 As can be seen from Fig.10(a), for $\alpha =10$, 
the frequency
 of small events are enhanced preceding the mainshock at and around the
 epicenter of the upcoming mainshock. For small enough $t$, such a
 cluster of smaller events correlated with the large event may be
 regarded as foreshocks. Just before the mainshock, the frequency of smaller
 events is suppressed in a close vicinity of the upcoming mainshock,
 while it continues to be enhanced in the surrounding blocks, a phenomenon closely resembling the ``Mogi doughnut'' \cite{Mogi69,Mogi79,Scholz02}. The spatial range where the quiescence occurs is narrow, only of a few  blocks. 

  For $\alpha=30$, as can be seen from Fig.10(b), the seismic
 acceleration preceding the mainshock is hardly discernible, while the
 doughnut-like quiescence is still realized. 
 We  note that the quiescence just before the mainshock is robustly observed in the BK model, independent of its dimensionality, the interaction range and 
 the parameter values. Indeed, it has been observed both in 1D and 2D, 
  both with the short-range and long-range interactions \cite{MK05,MK06}.

\subsection{SPATIAL CORRELATIONS OF EVENTS AFTER THE MAINSHOCK}

 In this subsection, we examine the time-development of spatial seismic
 correlations {\it after\/} the mainshock 
 of $\mu >\mu_c =5$. Figs.11(a) and (b) represent the cases of $\alpha=10$ in the  supercritical  regime and of $\alpha=30$ in the near-critical
 regime. Insets represent shorter-time  behaviors. Computational conditions are taken to be the same as in Figs.10.

\begin{figure}[ht]
\begin{center}
\includegraphics[scale=0.65]{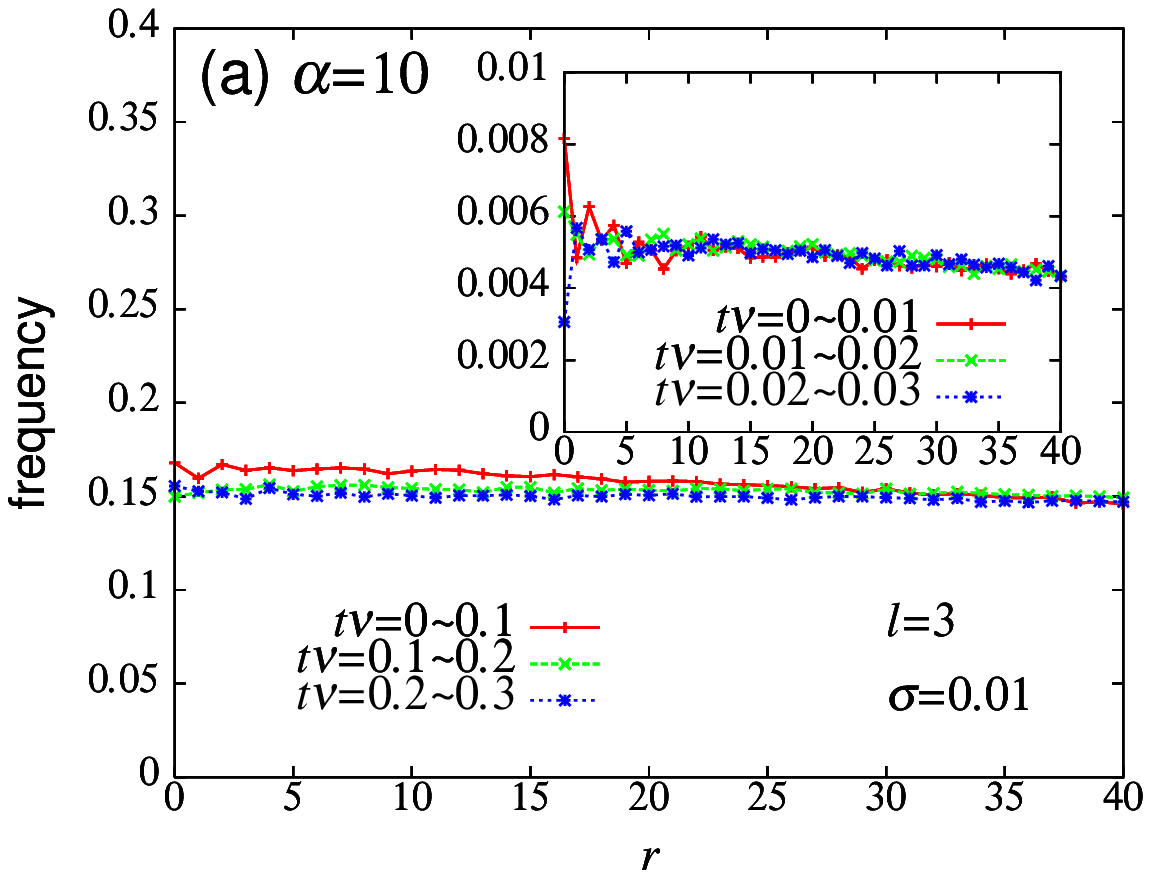}
\includegraphics[scale=0.65]{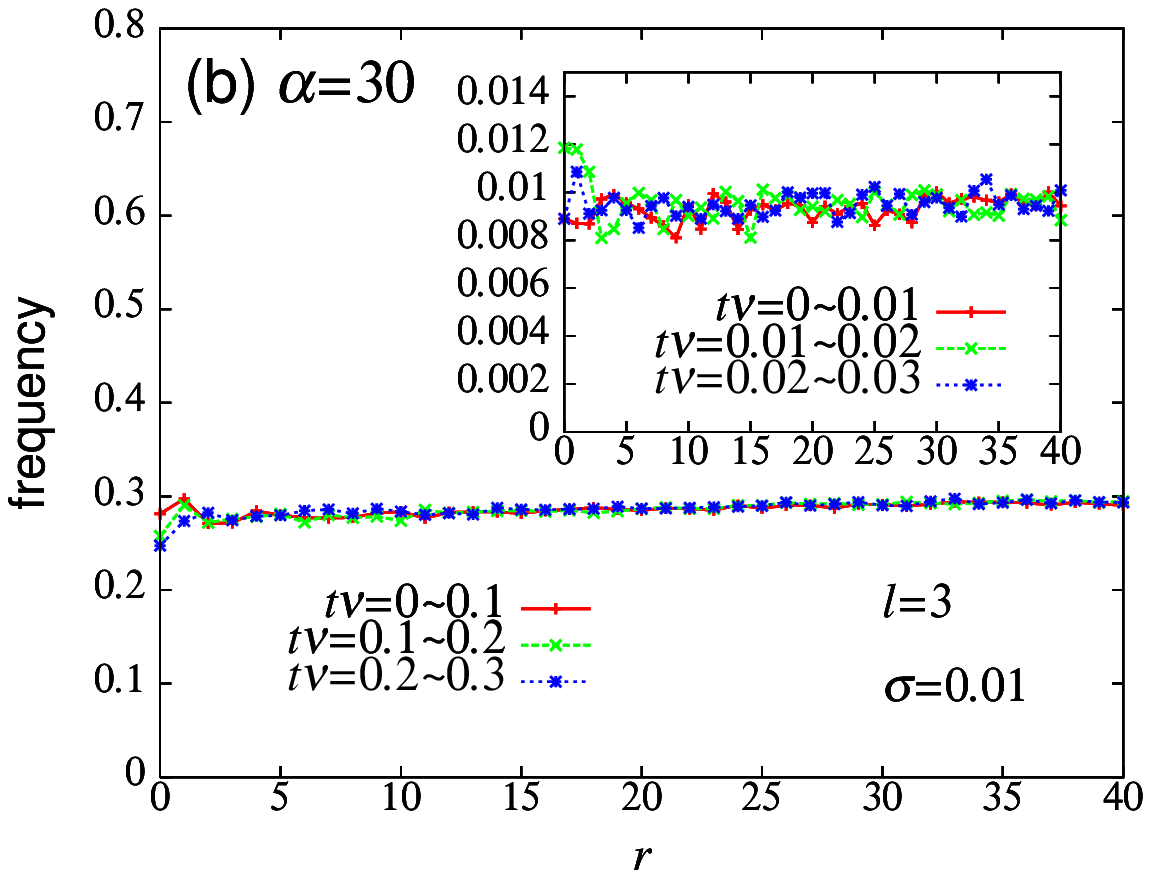}
\end{center}
\caption{
Event frequency after the large event of $\mu >\mu_c=5$
 (mainshock) plotted versus $r$,
the distance from the epicenter of the preceding mainshock, for several
 time periods after the mainshock in the 2D long-range BK model.
 The frictional-parameter $\alpha$ is 
$\alpha=10$ (a), and $\alpha=30$ (b) with $l=3$ and $\sigma=0.01$, each
 corresponding to the
 ``supercritical'' and ``near-critical'' regimes. The
 system size is $160\times 80$. The insets represent similar plots at
 shorter times.
}
\end{figure}

 As can be seen from the figures, spatiotemporal seismic correlations are almost absent after the mainshock in both cases of $\alpha=10$ and 
 $\alpha=30$. In the present 2D long-range model, the event  frequency hardly changes with distance $r$ nor with time $t$ even 
 in the supercritical regime, in contrast to the  short-range case where non-trivial spatiotemporal correlations are observed to some extent even after the mainshock \cite{MK08}.

\subsection{THE TIME-DEPENDENT MAGNITUDE DISTRIBUTION BEFORE THE MAINSHOCK}

\begin{figure}[ht]
\begin{center}
\includegraphics[scale=0.65]{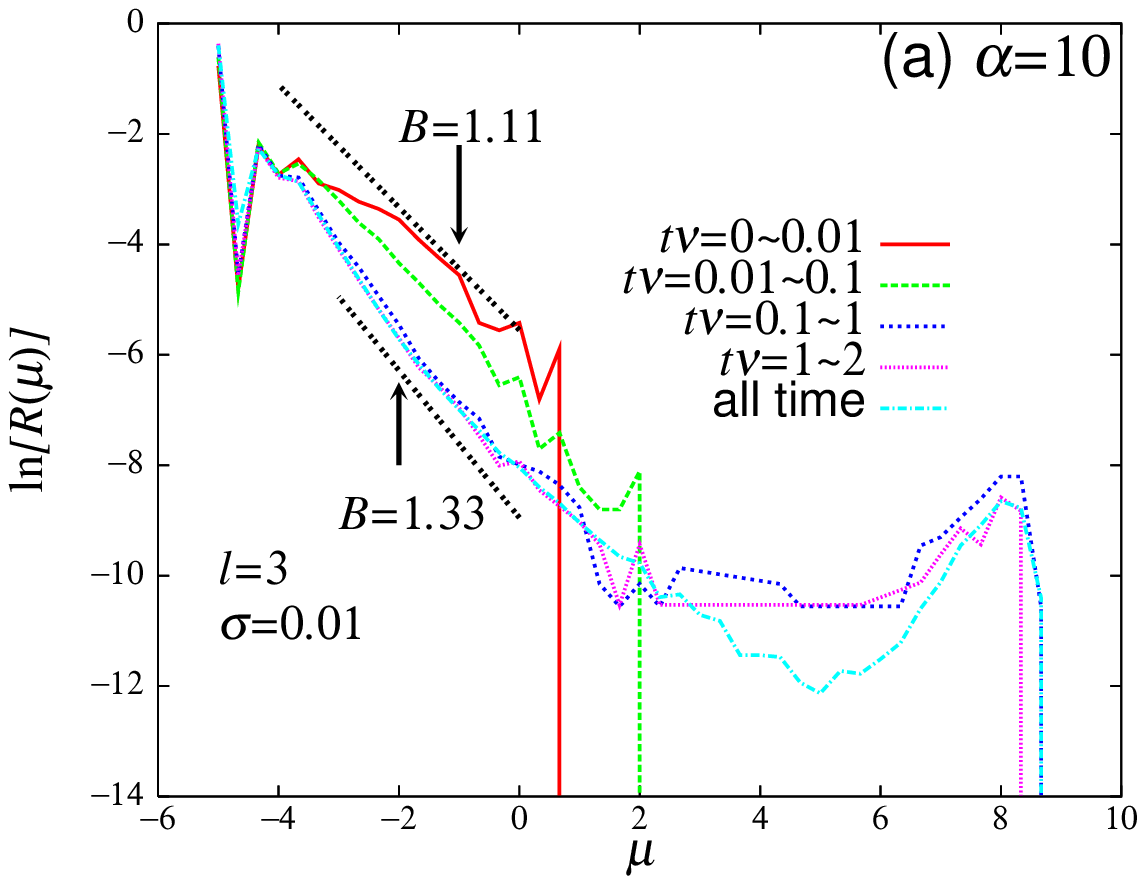}
\includegraphics[scale=0.65]{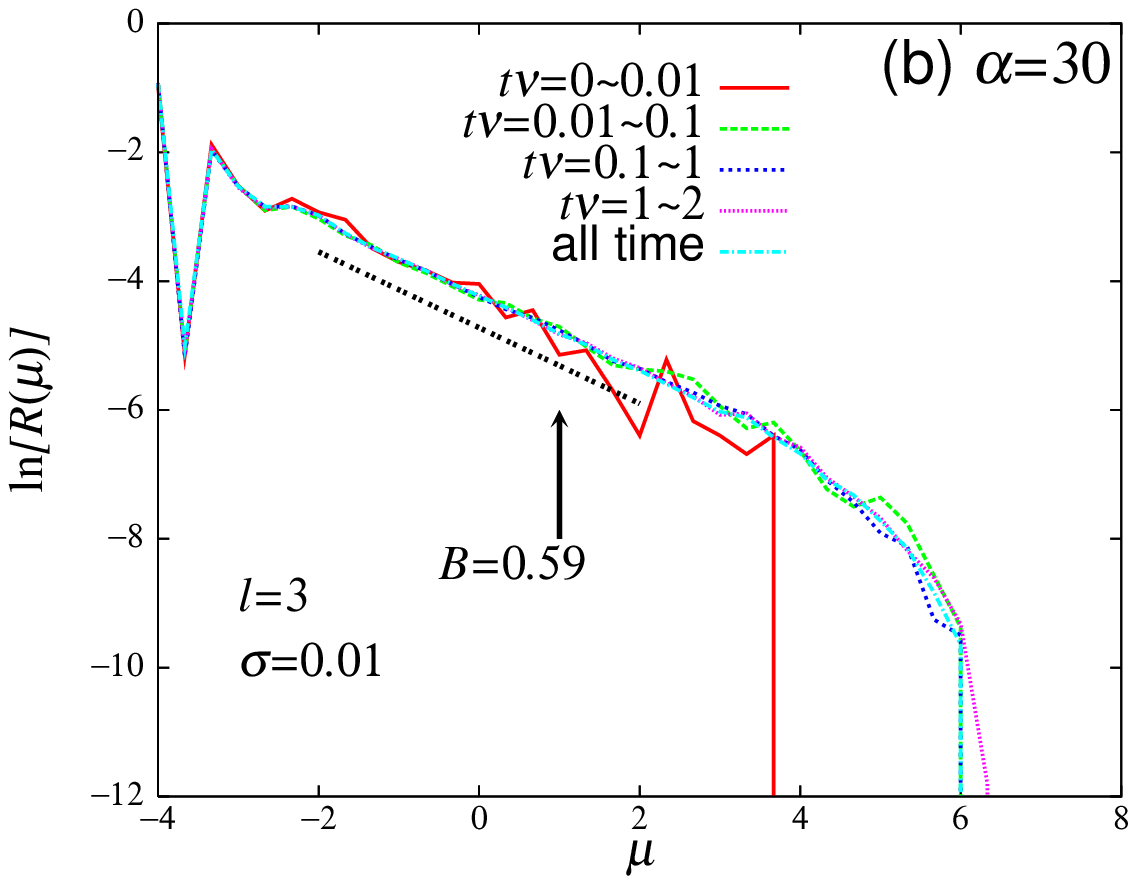}
\end{center}
\caption{
The local magnitude distribution of the 2D long-range BK model
 for several time periods before the mainshock of $\mu >\mu _c=3$ for the cases of $\alpha=10$ (a) and $\alpha=30$ (b), each corresponding
 to the ``supercritical'' and ``near-critical''
 regimes respectively. 
Events whose epicenter lies within 5 blocks from
 the epicenter of the upcoming mainshock are counted.
The parameters $l$ and $\sigma$ are fixed to $l=3$ and 
$\sigma =0.01$. The system size is $160\times 80$.  In (a), an
 apparent $B$-value decreases before the mainshock, while  in (b) it
 stays almost unchanged.
}
\end{figure}

  In real seismicity, an appreciable change of the $B$-value of the
 magnitude distribution has been reported preceding large earthquakes:
 Often a decrease of the $B$-value  \cite{Suyeetal64,JS99,Kawamura06}, 
 but sometimes
 an increase of it \cite{Smith81}. Obviously, a possible change in
 the magnitude distribution preceding the mainshock possesses a
 potential importance in earthquake prediction. 
 
In Figs.12, we show  the ``time-resolved''  local magnitude
 distributions for several time periods before the mainshock 
 for the cases of $\alpha=10$ (a) and of $\alpha =30$
 (b), with fixing $l=3$ and $\sigma=0.01$. 
 Only events with their epicenters lying within 5 blocks 
 from the upcoming mainshock  of $\mu\geq \mu_c=3$ are counted here. 

 As can be seen from Fig.12(a), in the supercritical regime, 
  an apparent $B$-value describing the
 smaller magnitude region, $\mu\lsim 2$, gets smaller as the mainshock
 is approached. Indeed, the $B$-value is reduced from the all-time value
  $B\simeq 1.33$ to $B\simeq 1.11$ here.  By contrast, as can   be seen from Fig.12(b),  an apparent
  $B$-value hardly changes in the near-critical regime even when the mainshock is approached. It   stays at around $B\simeq 0.59$. 

  Likewise, one can also study the ``time-resolved''  local magnitude  distributions {\it after\/} the large event. Seismic events are quite scarce after the large event, however, as can be seen from Figs.9 and 11, {\it i.e.\/}, few aftershocks observed. Hence, it is statistically difficult to obtain the ``time-resolved''  local magnitude  distributions {\it after\/} the mainshock in the  present model.

\section{SUMMARY AND DISCUSSION}

 Spatiotemporal correlations of the 2D BK model with the long-range interaction were studied by means of extensive numerical computer simulations.  The long-range interaction, which takes account of the effect of the elastic body adjacent to the fault plane, falls off with distance $r$ as $1/r^3$. Seismic properties of the model can be summarized in the form of a phase diagram shown in Fig.4. The 2D long-range model turns out to possess three distinct ``phases'', {\it i.e.\/}, a small-$\alpha$ near-critical phase, a supercritical phase and a large-$\alpha$ near-critical phase.

 The long-range BK model is certainly a more faithful representation of an earthquake fault than the short-range BK model.  Although some of the properties are more or less  common between in the short-range and in the long-range models, several  important differences exist in several observables.

 Generally speaking,  spatial seismic correlations of the long-range models tend to be suppressed compared with those of the short-range models. Such a suppression of spatial seismic correlations in the long-range model might intuitively be understandable, because the long-range nature of the interaction serves to smear out the spatial variation. 

 Most interestingly, it has been found that the magnitude distribution of the 2D long-range model exhibits a ``near-critical'' power-law-like behavior close to the Gutenberg-Richter law, for a wide parameter range with its $B$-value insensitive to the model parameter, $B\simeq 0.55$, in sharp contrast to the cases of the 2D short-range model and of the 1D short-range and long-range models where such  a near-critical behavior is realized only by fine-tuning the model parameter to a special value. Since the GR law is known to be robustly observed over different fault zones of varying locations and depths, possibly characterized by varying material parameters,  the power-law feature of the magnitude distribution should be a stable attribute of earthquake occurrence, not a special property requiring  a fine-tuning of the material parameter.  In that sense, stable occurrence of the near-critical magnitude distribution over a wide parameter range in the 2D long-range BK model might be of relevance to real seismicity. The observed $B$-value, $B\simeq 0.55$ ($b\simeq 0.83$), is not far from the one observed in real faults $B\simeq 2/3$ ($b\simeq 1$). It should be noticed that the near-critical magnitude distribution observed here is not a truly critical one, since, for sufficiently large magnitude, the magnitude distribution falls off sharply. The apparent power-law-like behavior does not extend toward larger magnitudes indefinitely.

  In real seismicity,  there holds an empirical law that the mean stress-drop of an earthquake  is nearly constant irrespective of the event magnitude. Although in the  short-range BK models, the mean stress-drop increases considerably as the  magnitude gets larger, in the long-range BK models the mean
 stress-drop hardly depends on the event magnitude in a wide parameter
 range. Hence, the long-range BK model in the near-critical regime has an obvious advantage that it can reproduce
 the observed constancy of the stress-drop. 

 Large events of the long-range model usually accompany foreshocks together with a doughnut-like quiescence as their precursors, while they hardly accompany aftershocks with almost negligible seismic correlations observed after the mainshock. Such absence of post-seismic activity correlated with the mainshock is more prominent in the long-range model than in the short-range model. Concerning pre-seismic activity preceding the mainshock, an appreciable change of the effective $B$-value has occasionally been observed both in the long-range and short-range models in 2D. The $B$-value is either increased, decreased or unchanged in 2D, depending on the system is in the subcritical, supercritical or near-critical phase.

 In this way, the long-range 2D BK model, which is apparently the most realistic version among the types of the BK model studied so far, appear to give a reasonable description of real seismicity in its near-critical regime. The model can explain the GR-like magnitude distribution with the $B$-value, $B\simeq 0.55$, stably realized over a rather wide parameter range, the near independency of the stress drop on the event magnitude and the absence of a characteristic peak in the recurrence-time distribution of earthquakes, {\it etc\/}. Meanwhile, characteristic features become most eminent in the supercritical regime, particularly for large events.

 Not all properties of real seismicity, however, are explained  by the model. For example, the Omori law frequently observed in real  seismicity cannot be reproduced in the BK model. This may suggest that the effects not taken account in the present model, {\it e.g.\/}, processes like the water migration through the crack, the slow chemical process at the fault or the elastoplasticity associated with the ascenosphere, are important in realizing the aftershock obeying the Omori law. 

 
In order to make a further link between the BK model and the
real world, we estimate here various time and
length scales involved in the BK model. For this, we need to estimate
the units of time and length of the BK model in terms of real-world
earthquakes. Concerning the time unit $\omega^{-1}$, we estimate it
via the rise time of large earthquakes, $\sim \pi / \omega$, which is
typically about 10 seconds. This gives an estimate of $\omega^{-1}
\sim 3$ sec. Concerning the length unit $\mathcal{L}$, we estimate it
making use of the fact that the typical displacement in large events
of our simulation is of order one $\mathcal{L}$ unit, which in
real-world large earthquakes is typically 5 meters. Then, we get
$\mathcal{L} \sim 5$ meters. Since the loading rate $\nu^{\prime}$
associated with the real plate motion is typically 5 cm/year, the
dimensionless loading rate 
$\nu=\nu^{\prime}/(\mathcal{L}\omega)$ is estimated to
be $\nu \sim 10^{-9}$. 

 In our simulation of the BK model, the doughnut-like quiescence was
observed before the mainshock at the time scale of, say, $t\nu \lsim
10^{-2}-10^{-1}$. This time scale corresponds to about 1-10 years. In our
simulation, the doughnut-like quiescence was observed in the region
only within a few blocks from the epicenter of the mainshock. To give
the corresponding real-world estimate, we need the real-world estimate
of our block size $a^{\prime}$. In the BK model, the length scale
$a^{\prime}$ is
entirely independent of the length scale $\mathcal{L}$, and has to be
determined independently. We estimate $a^{\prime}$ via the typical velocity of
the rupture propagation, $la^{\prime}\omega$, which is about 3 km/sec in real
earthquakes. From this relation, we get $a^{\prime} \sim 3$ km. The length
scale associated with the doughnut-like quiescence is then estimated
to be 3 $\sim$ 6 km in radius. 

 Throughout our present simulations, we have assumed the velocity-weakening friction force. The extent of the velocity weakening is mainly described by the parameter $\alpha$, which has a dimension of the inverse velocity. The unit of $\alpha^{-1}$ is then estimated to be $\sim 1$ m/sec. The recent high-velocity friction measurements indicate that the friction coefficient of serpentinite drops significantly at the slip velocity of order $0.1\sim 1$ m/sec \cite{Hirose}. This roughly corresponds to the $\alpha$-value of order $1\lsim \alpha \lsim 10$, which is indeed the values of interest here.

The seismic moment $M$ is approximately given by,
\begin{equation}
M \simeq GDS,
\end{equation}
where $G$ is the shear modulus, $D$ is 
the displacement and $S$ is the rupture-zone size. The shear modulus
$G$ of the crust is typically about 30 GPa. 
The moment magnitude $m$ measured in the MKS unit is defined by,
\begin{equation}
m=\frac{2}{3}\log_{10}M-6.
\end{equation}
From Eq.(7.1) and (7.2), we can obtain the relation between
the moment magnitude $m$ in real world and the magnitude $\mu$ in
the BK model as,
\begin{equation}
m\simeq 0.29\mu+6.09.
\end{equation}
In the BK model, an upper cut-off magnitude $\mu_{max}$ of the GR-like
behavior is found to be about 
$6 \lsim \mu_{max} \lsim 8$, which corresponds to $7.8 \lsim m_{max} \lsim
8.4$ in real world. Although it is tempting to speculate that the ``interrupted power-law'' or  ``near-criticality'' observed in our model simulation might somehow be related to real observation, it is not known whether there really exists an upper cut-off magnitude in real seismicity.

In the long-range BK model, the mean stress-drop hardly depends on the
event magnitude in the near-critical regime. 
In the 2D long-range BK model in the near-critical regime, the mean
stress-drop was estimated to be $\Delta \tau \simeq 1.6$. 
This corresponds in the real world the mean stress drop of 
$\simeq$ 80 MPa. Since the 
mean stress-drop is $1 \simeq 10$ MPa in real seismicity, the
estimated value of the mean stress-drop is a bit larger than but
roughly consistent with the real value.

 The present study was performed under many assumptions, {\it e.g.},
 an earthquake fault is completely flat, material parameters are
 homogeneous, there is no depth dependence in the material parameters, 
 a friction force 
 depends on the velocity alone, {\it etc.} As one of such 
 assumptions, we have employed a 
 static approximation in our simulation of the long-range BK model, 
 {\it i.e.}, we have assumed that  the velocity of the
 seismic-wave propagation is sufficiently larger than the rupture
 velocity.  In real earthquakes, however, the rupture velocity is
 comparable to the shear-wave velocity. Thus, it is clearly desirable
 to perform simulations  based on a fully dynamical elastic
 theory.

 In spite of such limitations of the model, our present study has revealed that the 2D long-range BK model can reproduce several important aspects of real seismicity. We hope that the present analysis might give a step toward the fuller understanding of the statistical properties of earthquakes.

\begin{acknowledgments}
 The authors are thankful to Prof. N. Kato and Dr. A. Ohmura for helpful discussion, and to Dr. T. Hasumi for correspondence.
\end{acknowledgments}

\appendix

\section{THE DERIVATION OF THE INTERACTION BETWEEN TWO ARBITRARY BLOCKS}

 In this appendix, based on an elastic theory, we derive the effective interaction between two arbitrary blocks on a fault plane in the 2D BK model. We begin with the static version of the representation theorem \cite{AR80} 
that represents a displacement in an elastic body induced
by a slip on a crack surface (or a fault plane) as
\begin{equation}
U_n(\Vec{x})=\int \!\!\! \int_{\Sigma^{\prime}} \Delta 
U_i(\Vec{x}^{\prime})C_{ijpq}n_j(\Vec{x}^{\prime}) \frac{\partial}
{\partial x_q^{\prime}}G_{np}(\Vec{x}
;\Vec{x}^{\prime})~d\Sigma^{\prime}, 
\end{equation}
where $U_n(\Vec{x})$ represents a displacement in the $n$-th direction at a spatial point $\Vec x$=($x_1,x_2,x_3$) in the elastic body,  
$\Delta U_i(\Vec{x}^{\prime})$ is a relative displacement in the
$i$-th direction across the fault surface $\Sigma$, $n_j(\Vec{x}^{\prime})$ 
is the normal unit vector on the fault surface, and $C_{ijpq}$ is an elastic constant. The Green's function 
$G_{np}(\Vec{x};\Vec{x}^{\prime})$ is a displacement in the
$n$-th direction at a  point $\Vec x$=($x_1,x_2,x_3$) due to a unit force acting along the $p$-th direction at a spatial point $\Vec x^{\prime}=(x_1^{\prime},x_2^{\prime},x_3^{\prime}$).
We assume the fault surface to be the $x_1 x_3$-plane which slips only in the $x_1$-direction. The elastic body is assumed to be isotropic,
homogeneous and infinite. 
 
 We consider a static version of the Navier's equation as a differential equation describing the elastic body,
\begin{equation}
(\lambda+\mu)\frac{\partial}
{\partial x_i}\left(\frac{\partial G_{jn}}{\partial x_j}\right)
+\mu\frac{\partial}
{\partial x_j}\left(\frac{\partial G_{in}}{\partial x_j}\right)=
-\delta_{in}\delta(x_1)\delta(x_2)\delta(x_3),
\end{equation}
the associated Green's function being given by 
\begin{equation}
G_{np}(\Vec{x};\Vec{x}^{\prime})=\frac{\delta_{np}}{4\pi\mu}\frac{1}{R}
-\frac{\lambda+\mu}
{8\pi\mu(\lambda+2\mu)}\frac{\partial^2 R}{\partial x_n\partial x_p}, 
\end{equation}
\begin{equation}
R=|\Vec x-\Vec x^{\prime}|=\sqrt{(x_1-x_1^{\prime})^2+(x_2-x_2^{\prime})^2+(x_3-x_3^{\prime})^2}, 
\end{equation}
where $\delta_{np}$ is the Kronecker's delta, and $\lambda$ and $\mu$ are 
Lame's constants. 

 The stress tensor $\tau _{ij}$ is related to the strain tensor $\epsilon_{ij}$ via the Hooke's law,
\begin{equation}
\tau_{ij} = \lambda \epsilon_{kk}\delta_{ij} + 2 \mu \epsilon_{ij}, 
\end{equation} 
\begin{equation}
\epsilon_{kl} \equiv \frac{1}{2}\left(\frac{\partial U_k}{\partial x_l} +
\frac{\partial U_l}{\partial x_k} \right). 
\end{equation}
 Then, we consider the situation where an infinitesimal part of the fault plane  $d\Sigma^{\prime}$ located at $(x_1^{\prime},0,x_3^{\prime})$ slips by an amount $\Delta U$. By using Eqs.(A1), (A3), one gets the stress on the $x_1x_3$-plane ($x_2=x_2^{\prime}$) as
\begin{equation}
\begin{array}{ll}
\tau_{12}(x_1-x_1^{\prime},0,x_3-x_3^{\prime})=
\ \ \ \ \\ \ \ \ \ \frac{\mu(3\lambda+2\mu)d\Sigma}
{4\pi(\lambda+2\mu)}\left[\frac{(x_1-x_1^{\prime})^2}{R_0^5}+\frac{2\mu}
{3\lambda+2\mu}\frac{(x_3-x_3^{\prime})^2}{R_0^5}
\right]\Delta U, 
\end{array}
\end{equation}
\begin{equation}
R_0=\sqrt{(x_1-x_1^{\prime})^2+(x_3-x_3^{\prime})^2}.
\end{equation}
The result means that the stress decays with distance $R_0$ as $1/R_0^3$ on the fault plane. Indeed, Maruyama discussed a static version of the three-dimensional source mechanics of earthquakes \cite{Maru64}. 
The result we have obtained here corresponds to his result. 

 Now, we wish to apply Eqs. (A7) and (A8) derived from an elastic theory to the BK model. First, we discretize the fault plane into  blocks of linear size $a^{\prime}$. Second, we regard $\Delta U$ to be a {\it relative\/} displacement between two blocks, {\it i.e.\/}, $\Delta U=U_{i,j}-U_{i^{\prime},j^{\prime}}$, where $U_{i,j}$ denotes a displacement of a block at a site (${i,j}$). 
Note that, by this choice of $\Delta U$,  one has a vanishing self-interaction, since the relative displacement with itself always vanishes.


The spring constant $K(i,j;i^{\prime},j^{\prime})(U_{i,j}-U_{i^{\prime},j^{\prime}})\equiv {a^{\prime}}^2\tau$ 
between the blocks at a site (${i,j}$) and at (${i^{\prime},j^{\prime}}$) is then given by
\begin{equation}
K(i,j;i^{\prime},j^{\prime})={l_x^{\prime}}^2\frac{(i-i^{\prime})^2}{r^5}+{l_z^{\prime}}^2\frac{(j-j^{\prime})^2}{r^5}, 
\end{equation}

\begin{equation}
{l_x^{\prime}}^2=\frac{\mu(3\lambda+2\mu)a^{\prime}}{4\pi(\lambda+2\mu)},
\end{equation}
\begin{equation}
{l_z^{\prime}}^2=\frac{2\mu^2a^{\prime}}{4\pi(\lambda+2\mu)}, 
\end{equation}
\begin{equation}
r=\sqrt{(i-i^{\prime})^2+(j-j^{\prime})^2}.
\end{equation}

The dimensionless inter-block interaction is then given by

\begin{equation}
({l_x}^2\frac{(i-i^{\prime})^2}{r^5}+{l_z}^2\frac{(j-j^{\prime})^2}{r^5})(u_{i,j}-u_{i^{\prime},j^{\prime}}), 
\end{equation}

\begin{equation}
{l_x}^2=\frac{{l_x^{\prime}}^2}{k_p}=\frac{\mu(3\lambda+2\mu)a^{\prime}}{4\pi(\lambda+2\mu)k_p},\ \ 
\end{equation}
\begin{equation}
{l_z}^2=\frac{{l_z^{\prime}}^2}{k_p}=\frac{2\mu^2a^{\prime}}{4\pi(\lambda+2\mu)k_p}, 
\end{equation}
where $k_p$ is the spring constant introduced in \S 3, and $u_{i,j}$ is a dimensionless displacement defined in \S 3.

 Similarly, for the long-range 1D BK model, 
we can obtain the interaction
$K(i;i^{\prime})$ between two arbitrary blocks at site $i$ and 
$i^{\prime}$. In the long-range 1D BK model, 
we have assumed the fault and the elastic body to be a rigid body in the $x_3$-direction. 
 In this case, a static version of the Navier's equation may be 
written as
\begin{equation}
(\lambda+\mu)\frac{\partial}
{\partial x_i}\left(\frac{\partial G_{jn}}{\partial x_j}\right)
+\mu\frac{\partial}
{\partial x_j}\left(\frac{\partial G_{in}}{\partial x_j}\right)=
-\delta_{in}\delta(x_1)\delta(x_2),
\end{equation}
the associated Green's function being given by
\begin{equation}
G_{np}(\Vec{x};\Vec{x}^{\prime})=-\frac{\delta_{np}}{2\pi\mu}\log{R}
+\frac{\lambda+\mu}{8\pi\mu(\lambda+2\mu)}
\frac{\partial^2 [R^2(\log{R}-1)]}{\partial x_n\partial x_p}, 
\end{equation}
\begin{equation}
R=|\Vec x-\Vec x^{\prime}|=\sqrt{(x_1-x_1^{\prime})^2+(x_2-x_2^{\prime})^2}. 
\end{equation}
By using Eqs.(A1) and (A13), 
one obtains the stress on the $x_1x_3$-plane ($x_2=x_2^{\prime}$) as
\begin{equation}
\tau_{12}(x_1-x_1^{\prime},0)=
\frac{\mu(\lambda+\mu)d\Sigma}
{\pi(\lambda+2\mu)} \frac{1}{(x_1-x_1^{\prime})^2} \Delta U. 
\end{equation}
Then, after the block discretization and the replacement $\Delta U=U_{i}-U_{i^{\prime}}$, the spring constant defined by $K(i;i^{\prime})(U_{i}-U_{i^{\prime}})\equiv a^{\prime}\tau$ is obtained as
\begin{equation}
K(i;i^{\prime})=
{l^{\prime}}^2\frac{1}{|i-i^{\prime}|^2}, 
\end{equation}

\begin{equation}
{l^{\prime}}^2=
\frac{\mu(\lambda+\mu)a^{\prime}}
{\pi(\lambda+2\mu)}. 
\end{equation}

The dimensionless inter-block interaction is then given by

\begin{equation}
l^2\frac{1}{|i-i^{\prime}|^2}(u_{i}-u_{i^{\prime}}), 
\end{equation}

\begin{equation}
{l}^2=\frac{{l^\prime}^2}{k_p}=
\frac{\mu(\lambda+\mu)a^{\prime}}
{\pi(\lambda+2\mu)k_p}. 
\end{equation}

\section{THE 1D BK MODEL WITH THE LONG-RANGE INTERACTION}

In this appendix, we show some of the results of our numerical
simulations on the 1D BK model with the long-range interaction decaying as
$1/r^2$.

\begin{figure}[ht]
\begin{center}
\includegraphics[scale=0.65]{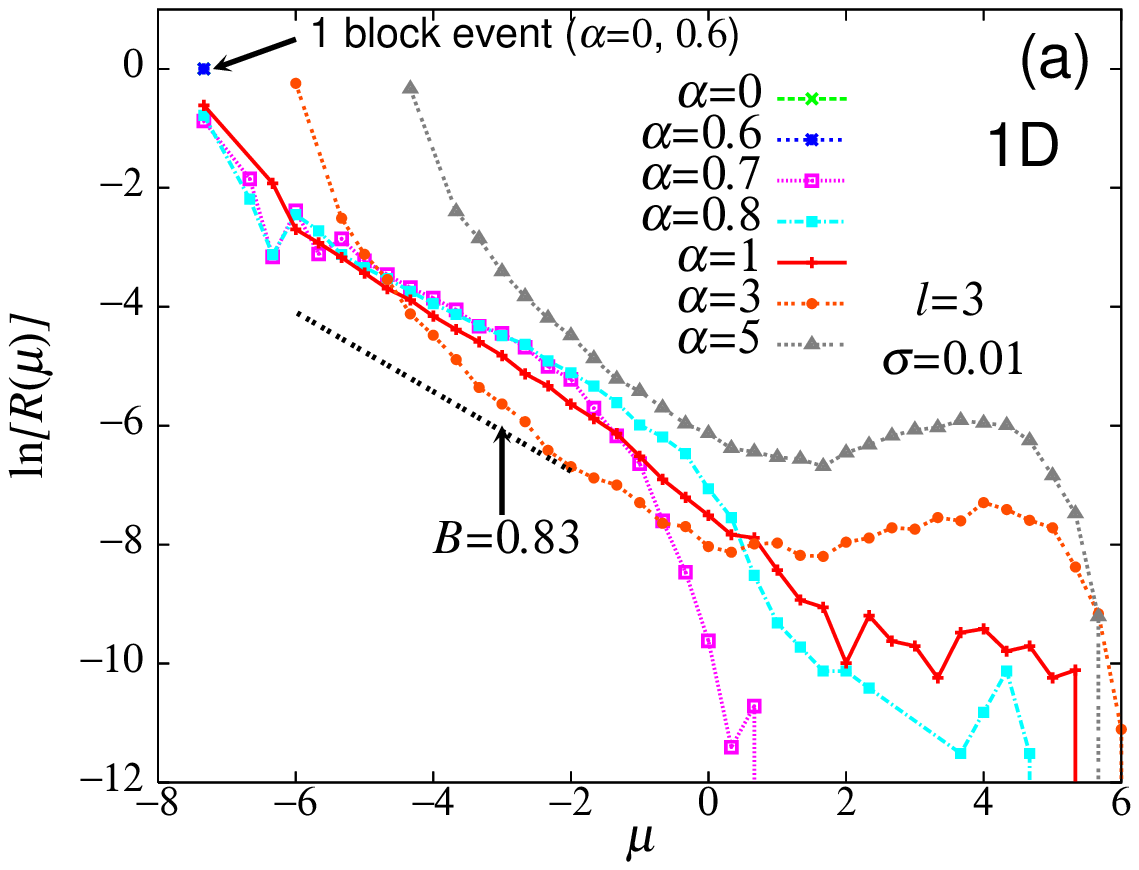}
\includegraphics[scale=0.65]{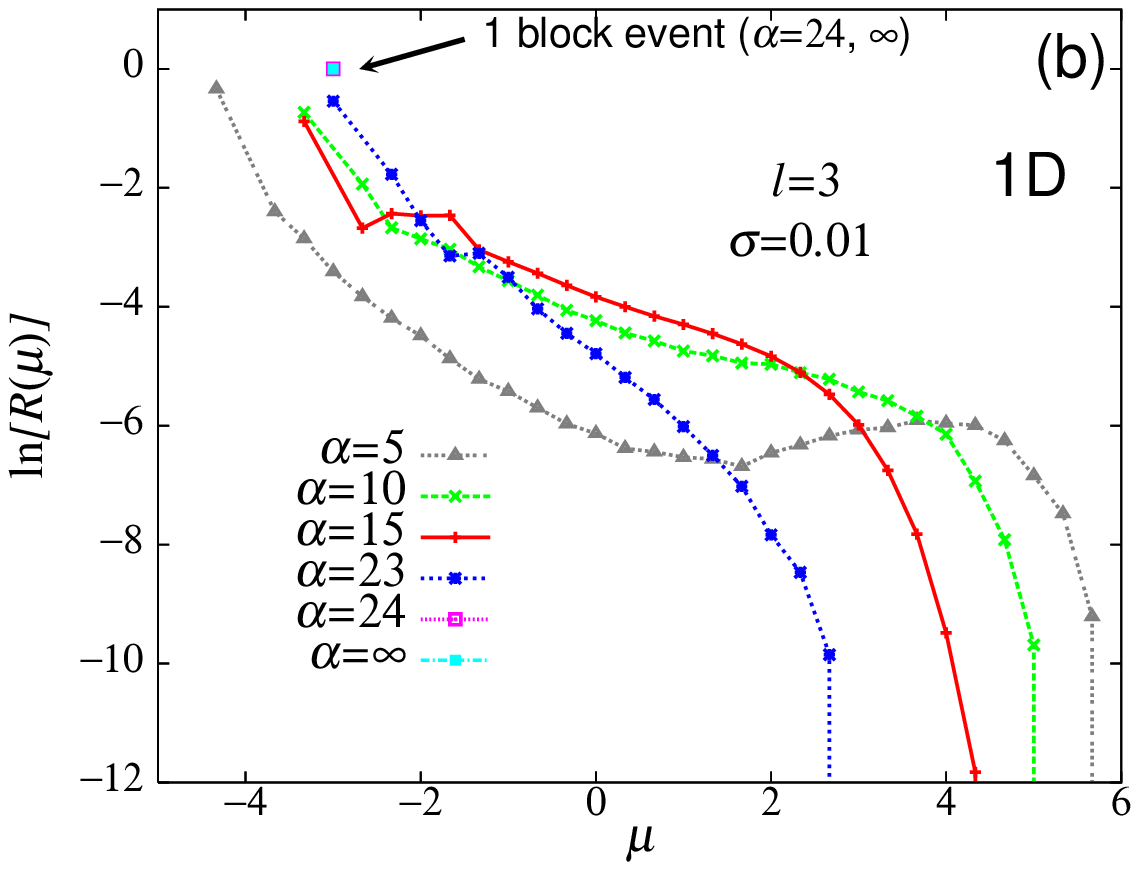}
\end{center}
\caption{
The magnitude distribution $R(\mu)$ of earthquake events of the 
 1D long-range BK model  for the  parameters $l=3$ and $\sigma=0.01$. Fig.(a) represents
 $R(\mu)$ for  smaller values of the frictional parameter 
 $0\leq \alpha \leq 5$, while Fig.(b) represents $R(\mu)$ for larger values of
 the frictional parameter $5\leq \alpha \leq \infty$. The system size is
 $N=800$.
}
\end{figure}

 Typical behaviors of the magnitude distribution are  shown in Figs.1 for the case of $l=3$ and $\sigma=0.01$.   Figs.13(a) and (b) exhibit $R(\mu)$ for smaller and larger $\alpha$,  respectively. The peculiarity of the 1D long-range BK model is that, for sufficiently small values of $\alpha \lsim 0.6$,  only one-block events occur under periodic boundary condition in the steady state realized after transients. Under free boundary condition, on the other hand, such an exclusive occurrence of one-block events does not arise for any $\alpha$. We note here  that the behavior for smaller $\alpha$ is rather sensitive to the choice of the time discretization $\Delta t$. In the region of smaller $\alpha$, we need to take $\Delta t$ as small as $10^{-5}$ to get stable results. Otherwise, totally different behaviors would sometimes arise. 

 In the range of $0.7 \lsim \alpha \lsim 1$, events involving more than one block begin to occur, where the associated  $R(\mu)$ exhibits a  ``subcritical'' behavior bending down rapidly at larger magnitudes, as can be seen from Fig.13(a). As $\alpha$ is increased, weights of larger events tend to increase gradually, and at $\alpha =1$, $R(\mu)$ exhibits a near straight-line  ``near-critical'' behavior close to the GR-law behavior. 

 As $\alpha$ is increased further beyond $\alpha=1$,  $R(\mu)$ develops a characteristic peak and exhibits a ``supercritical'' behavior,  deviating from the GR law at larger magnitudes $\mu\gsim \tilde \mu\simeq 1$,  while it still exhibits a near straight-line behavior corresponding to the GR law at smaller magnitudes $\mu\lsim \tilde \mu$.  As $\alpha$ is further increased, the peak at a larger magnitude becomes less pronounced,  and at $\alpha \simeq 15$, $R(\mu)$ exhibits a near-critical behavior again without a characteristic peak. For $\alpha \gsim 15$, $R(\mu)$ exhibits a subcritical behavior, rapidly bending down at larger magnitudes.  Finally, events involving more than one block suddenly disappear. In the range of $\alpha \gsim 24$,  only one-block events occur. As in the case of smaller $\alpha$,  we need to take  the time discretization $\Delta t$ sufficiently small in order to correctly reproduce such a behavior in this regime of larger $\alpha$.
 
 While the magnitude distributions presented here are the first data on the 1D BK model with the $1/r^3$ long-range interaction, we wish to make some comparison with the earlier data for the related 1D BK models. The magnitude distribution of the 1D short-range (nearest-neighbor) BK model was studied by several authors, including the earlier calculation of Carlson, Langer and collaborators \cite{CL89b,Carletal91} as well as of our own \cite{MK08}. The data of Ref.\cite{CL89b,Carletal91} corresponded to the ``supercritical'' regime  ($\alpha=2.5,3$ and 4) and the ``near-critical'' regime ($\alpha=1$). Our present data are qualitative similar to those of Refs.\cite{CL89b,Carletal91} in these regimes, though the GR-like behavior at smaller magnitudes, {\it i.e.\/}, the linearity of the $R(\mu)$ curve, seems less pronounced in our present case and in Ref.\cite{MK08} than in Ref.\cite{CL89b,Carletal91}. This is due to the different choice of the $l$-value: Carlson {\it et al\/} took $l$ to be large $6\sim 14$, while we mostly choose $l=3$ here and in Ref.\cite{MK08}.

 By contrast, if we compare our present $R(\mu)$ for the  $1/r^3$ long-range BK model with the one obtained  in Ref.\cite{Xiaetal05,Xiaetal07} for the mean-field-type long-range BK model, there exists some appreciable qualitative difference. Namely, even in the ``supercritical'' regime of $\alpha=2$ and 2.5, the magnitude distribution of Ref.\cite{Xiaetal05,Xiaetal07} exhibits no characteristic peak at a larger magnitude, but rather exhibits a down-bending ``subcritical''-type behavior. In Ref.\cite{Xiaetal05, Xiaetal07}, a characteristic peak in  $R(\mu)$ is discernible in the region of smaller $\alpha$ ($\alpha=0, 0.5$) where we have observed here either ``one-block events only'' behavior or ``subcritical'' behavior without a characteristic peak. We have checked that this qualitative difference is not due to the different choice of the $l$-value in the two calculations. Thus, the behavior of the magnitude distribution appears to differ substantially between in the mean-field-type long-range model and in the  $1/r^3$ long-range model.

\begin{figure}[ht]
\begin{center}
\includegraphics[scale=0.65]{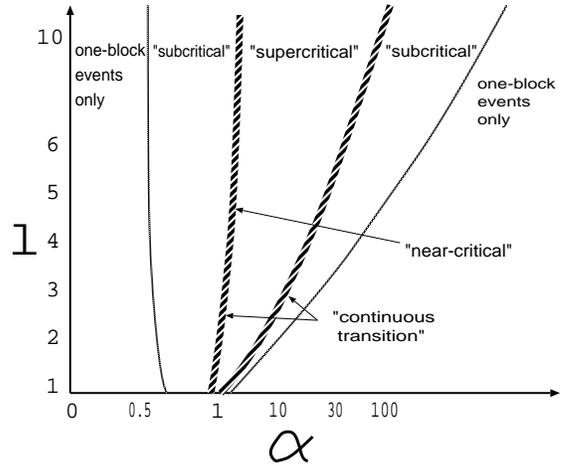}
\end{center}
\caption{
The phase diagram of the 1D BK models with the long-range
 interaction in the frictional-parameter $\alpha$ versus
 elastic-parameter $l$ plane. The parameter $\sigma$ is  set to $\sigma=0.01$.  To draw a phase diagram, the parameter range $0\leq \alpha \leq \infty$ and $1\leq l \leq 10$ is studied by simulations.
}
\end{figure}

 In Fig.14, we summarize the behavior of $R(\mu)$ of the 1D long-range BK model in the form of a
 ``phase diagram'' in the frictional-parameter $\alpha$ versus the
 elastic-parameter $l$ plane for the case of $\sigma=0.01$. The phase
 diagram consists of five distinct regimes,  two of which are ``one-block events'' regimes, two are 
 ``subcritical'' regimes and one is a ``supercritical'' regime. The
transition between  the small-$\alpha$ subcritical regime and the supercritical regime appears to be continuous (gradual), in contrast to the one of the 2D long-range model. The transition between  different ``phases'' is primarily dictated  by the $\alpha$-value. Since the ``phase boundary'' in Fig.14
 has a finite slope in the $\alpha$-$l$ plane,  one can also induce the
 transition by increasing the $l$-value for a fixed $\alpha$.


 In the mains panels of  Figs.15(a)-(c), we  show the magnitude dependence of the mean displacement, the mean number of  failed-blocks and the mean stress-drop of the 1D long-range BK model for various values of  $\alpha$. In the insets, we show the system-size dependence of each quantity for the case of $\alpha=1$.

\begin{figure}[ht]
\begin{center}
\includegraphics[scale=0.6]{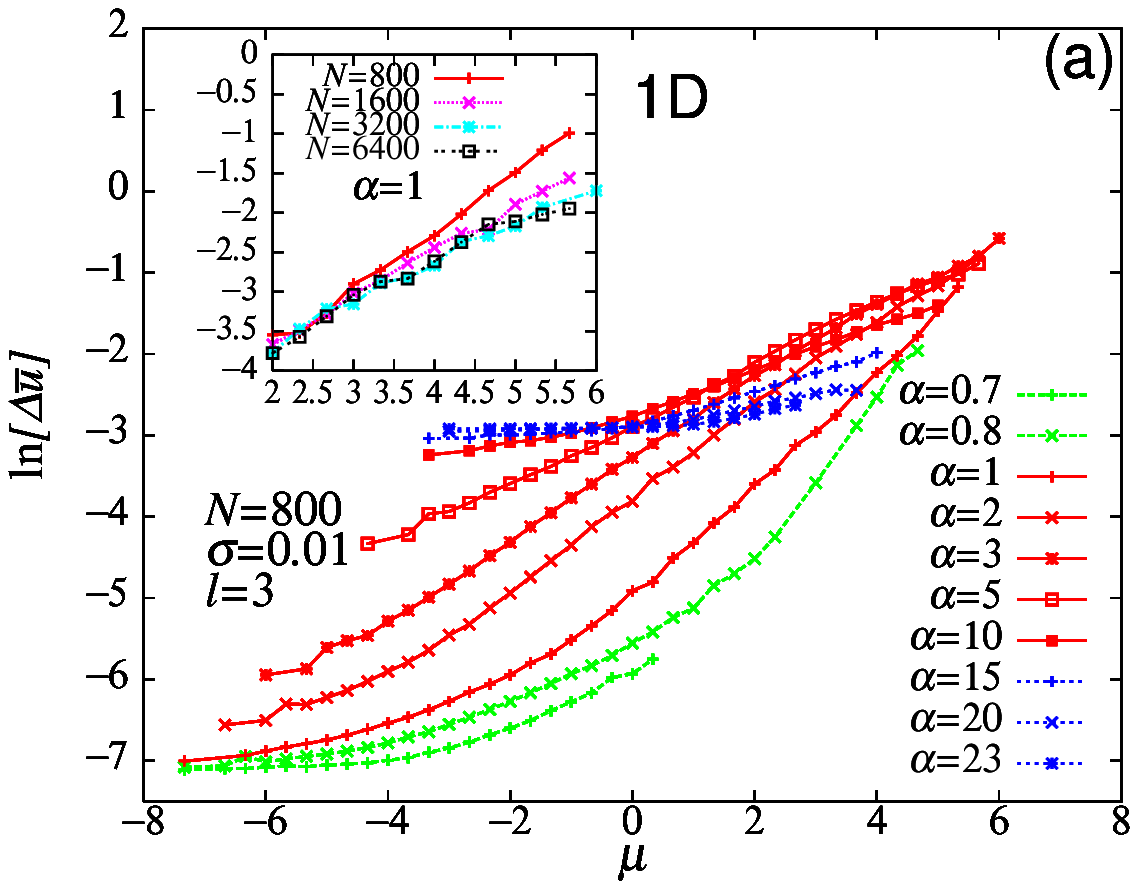}
\includegraphics[scale=0.6]{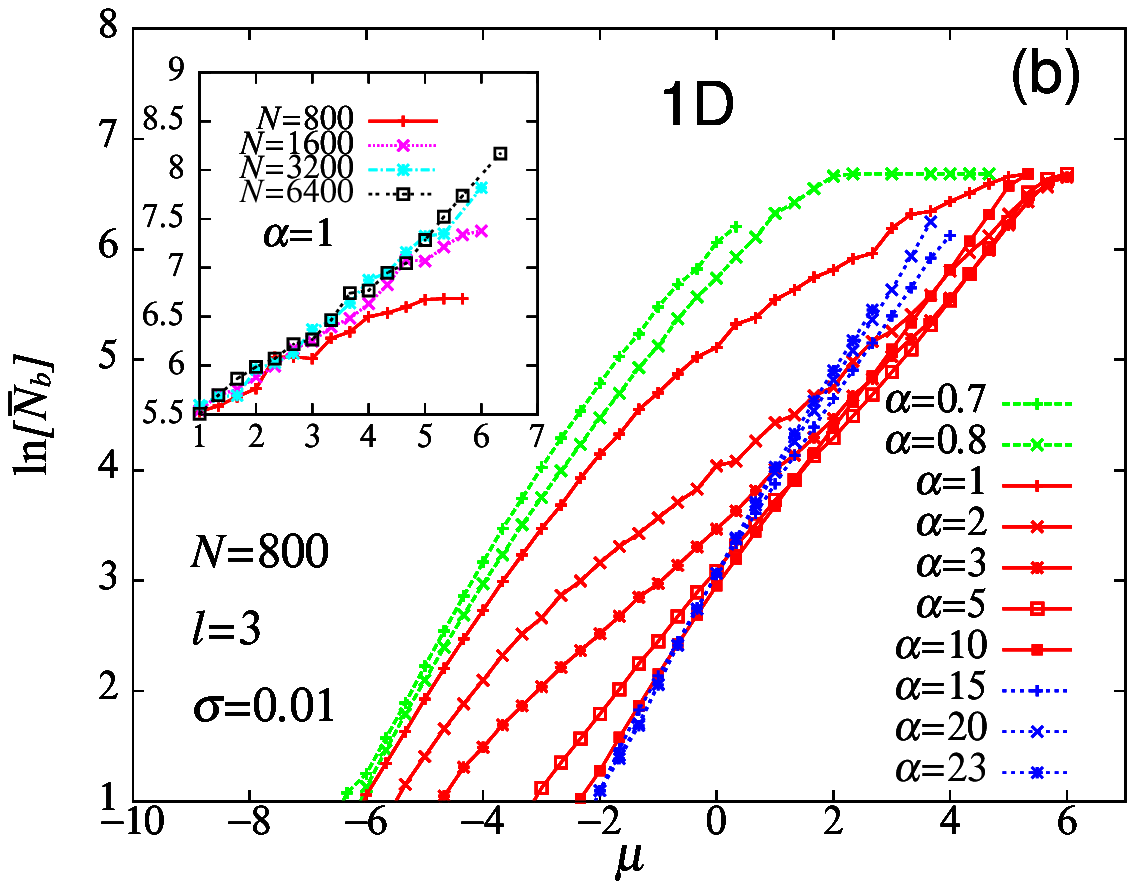}
\includegraphics[scale=0.6]{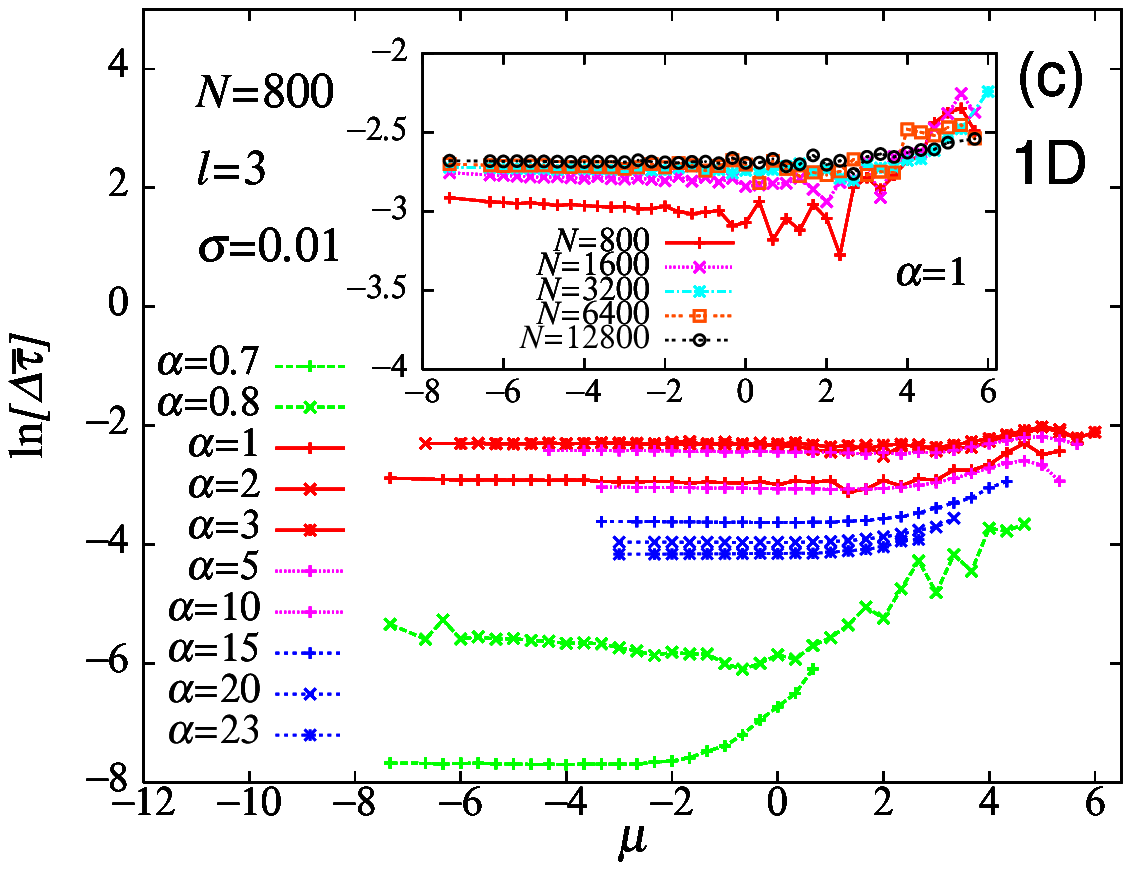}
\end{center}
\caption{
 The magnitude dependence of the mean displacement (a),
 the mean number of failed-blocks (b), and the mean stress-drop (c) of each seismic event of the 1D long-range BK model. In the main panels, the frictional-parameter $\alpha$ is varied with fixing the system-size $N=800$, while in the insets the system-size $N$ is varied for $\alpha=1$. 
  The parameters $l$ and
 $\sigma$ are fixed to $l=3$ and $\sigma=0.01$.
}
\end{figure}

 As can be seen from Figs.15(a) and (b), the data might roughly be grouped into three different categories, each corresponding to the small-$\alpha $ subcritical regime, the supercritical regime and the large-$\alpha$ subcritical regime, although the transition between these behaviors is rather gradual. As compared with the corresponding 2D models, including both the short-range model studied in \cite{MK08} and the long-range model studied in \S 3, the scaling property is much more obscured here in 1D. The data in the subcritical regimes do not collapse on top of each other, nor exhibit a straight-line power-law-like  behavior.

 As can be seen from Fig.15(c), the mean stress-drop of a seismic event  $\Delta \bar{\tau}$ hardly depends on its magnitude  $\mu$ except for large earthquakes. There is even a tendency that the mean stress-drop becomes more independent of the event magnitude as one studies larger systems (see the inset). Similar independence is also observed in the 2D long-range model in \S 3, as well as in Ref.\cite{Xiaetal07} for the mean-field-type 1D long-range model, and might be contrasted to the property of the corresponding short-range model where the mean stress-drop exhibits more pronounced magnitude dependence \cite{MK08}.

\begin{figure}[ht]
\begin{center}
\includegraphics[scale=0.58]{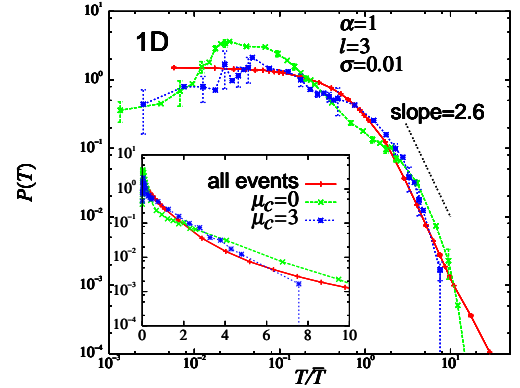}
\end{center}
\caption{
 The local recurrence-time distribution function $P(T)$ of the 1D long-range 
 BK model of the frictional parameter $\alpha=1$, with
 varying the magnitude threshold $\mu_c$. 
 The main panels represent the log-log plot of $P(T)$ and 
 the insets represent the semi-logarithmic plots including the 
 tail part of the distribution. The mean recurrence time $\bar T$ is
 $\bar T=0.000916, 0.369$ and 19.7, respectively for $\mu_c=-\infty, 0$ and 3.
}
\end{figure}

 We have also computed the local recurrence-time distribution $P(T)$ for events  of their  magnitude $\mu \geq \mu _c$.  The local recurrence time $T$ is defined by the time passed until the next event occurs with its epicenter lying in a  vicinity of the previous event within distance of $r=30$-blocks from the epicenter of the previous event. The behavior of the computed local recurrence-time distribution is qualitatively similar to the one of the 2D long-range model given in \S 3; an exponential tail at longer $T$, with or without a characteristic peak at shorter $T$ in the supercritical or in the subcritical regimes, respectively. 

 One case of interest in the 1D model might be the $\alpha=1$  near-critical case located at the phase boundary between the small-$\alpha$ subcritical regime and the supercritical regime, since such a region is absent in the corresponding 2D model due to the discontinuous nature of the transition. Thus, in Fig.16, we show on a log-log plot the computed $P(T)$ for the case of $\alpha=1$, with fixing $l=3$ and $\sigma=0.01$, for various values of the magnitude threshold $\mu_c$. As can be seen from the figure, $P(T)$ tends to exhibit a power-law-like behavior at larger $T$ as the magnitude threshold $\mu_c$ is taken smaller. The associated exponent is estimated to be about $\simeq 2.6$. This suggests that,  at $\alpha=1$, the occurrence of small events has a critical feature, while such a critical feature is weakened for larger events. Such a critical feature was not seen in the recurrence-time distribution of the 2D long-range model studied in \S 3. We note that, even in 1D, such a critical $P(T)$ is realized only at $\alpha=1$. For other values of $\alpha$, $P(T)$ robustly exhibits an exponential tail at longer $T$ (not shown here).

 We have also calculated various spatiotemporal correlation functions for the 1D long-range BK model, most of which show behaviors qualitatively similar to the ones observed for the 2D long-range BK model. Among them, we show in Figs.17 the  ``time-resolved''  local magnitude
 distributions for several time periods before the large event
 for the cases of $\alpha =1$ (a), $\alpha=5$ (b) and $\alpha=15$ (c), with fixing $l=3$ and $\sigma=0.01$.
 Only events with their epicenters lying within 30 blocks
 from the upcoming mainshock is counted here. We define the mainshock as
 a large event of $\mu\geq \mu_c=3$. 

 For the case of $\alpha=1$, as shown in Fig17(a),  an apparent $B$-value describing the  smaller magnitude region $\mu\lsim -1$ gets smaller from the all-time value $B\simeq 0.79$ to the short-time value $B\simeq 0.67$ as the mainshock
 is approached. Such a decrease of the $B$-value is opposite to the one observed in the corresponding 1D short-range model at $\alpha=1$ where the $B$-value  gets larger as the mainshock is approached \cite{MK05,MK06}.

 For the case of $\alpha=15$, by contrast, 
 an apparent $B$-value describing the smaller magnitude region gets larger as the mainshock is approached: See Fig.17(c).
  
 For the case of $\alpha=5$, the time
 development of the magnitude distribution exhibits a somewhat different behavior as shown in Fig.17(b). As  the mainshock is approached, the magnitude distribution $R(\mu)$  is developed from the supercritical all-time behavior to the near-critical straight-line behavior characterized by a slope $B\simeq 1.11$.

\begin{figure}[ht]
\begin{center}
\includegraphics[scale=0.6]{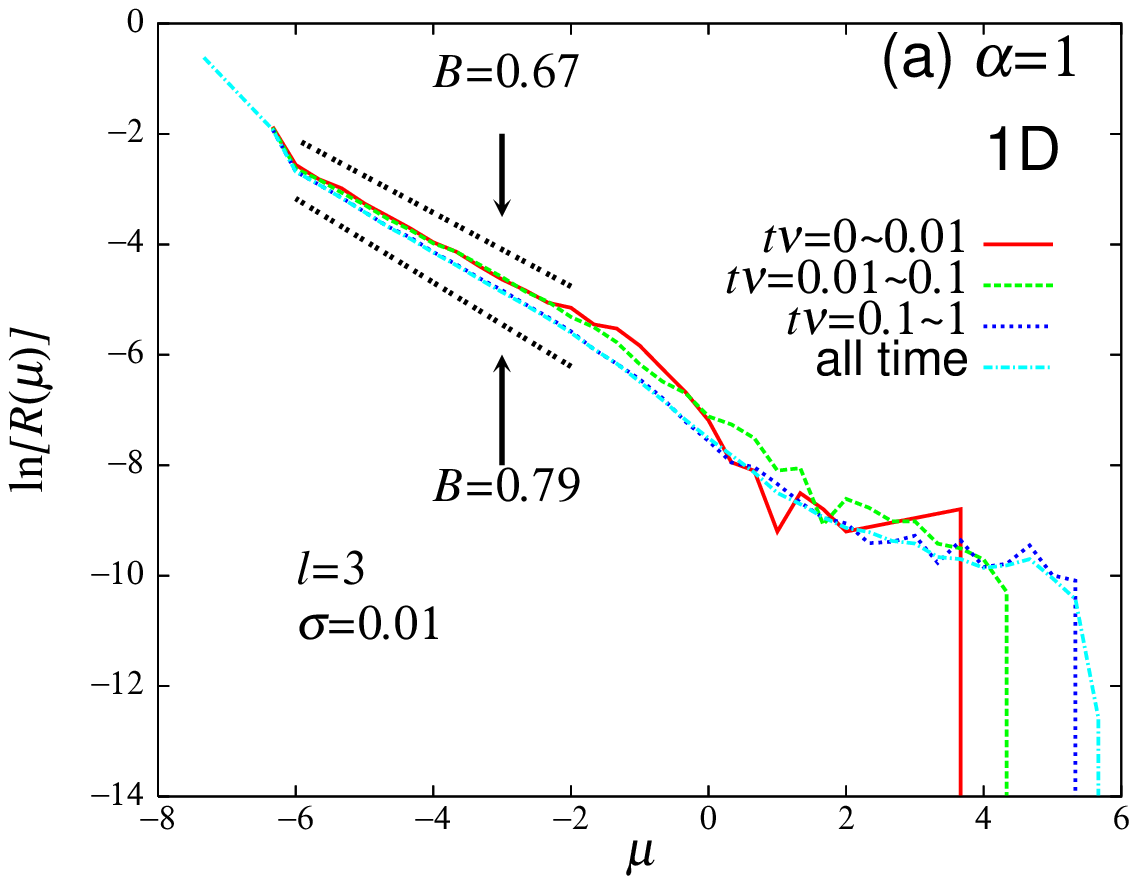}
\includegraphics[scale=0.6]{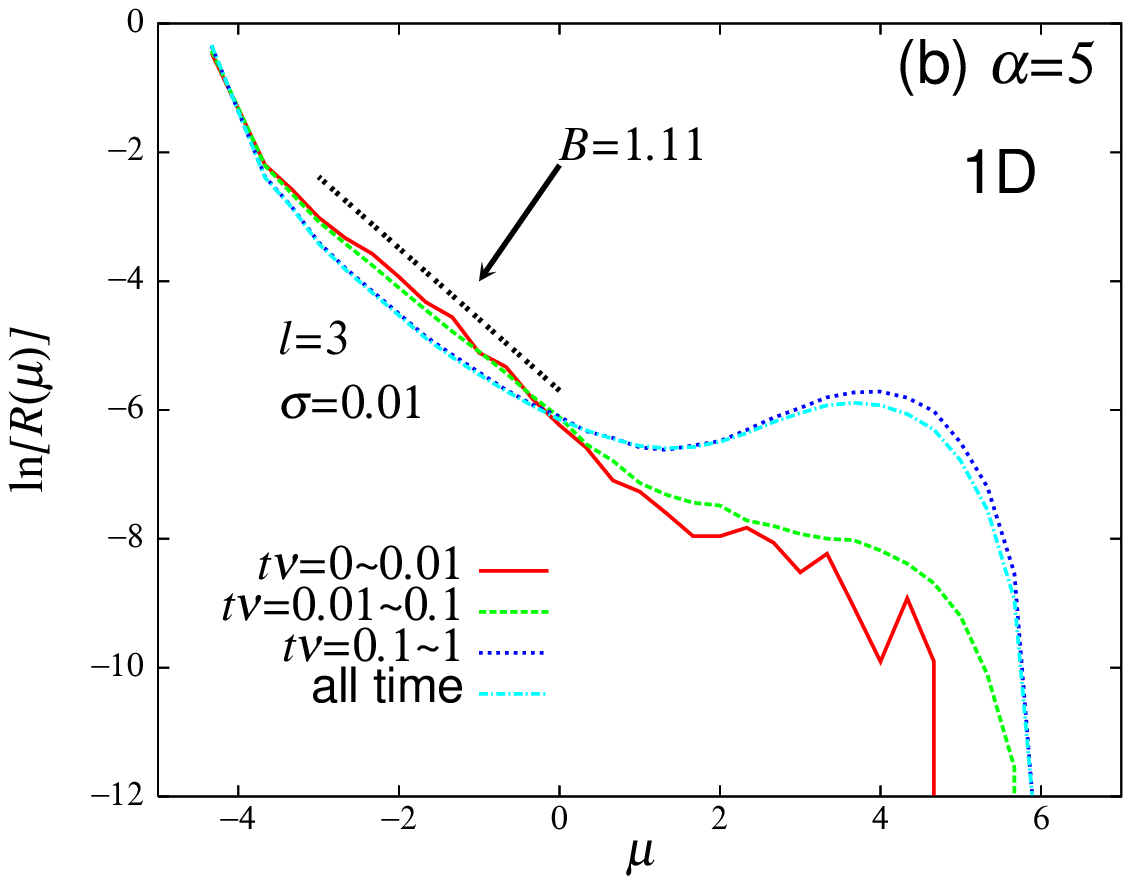}
\includegraphics[scale=0.6]{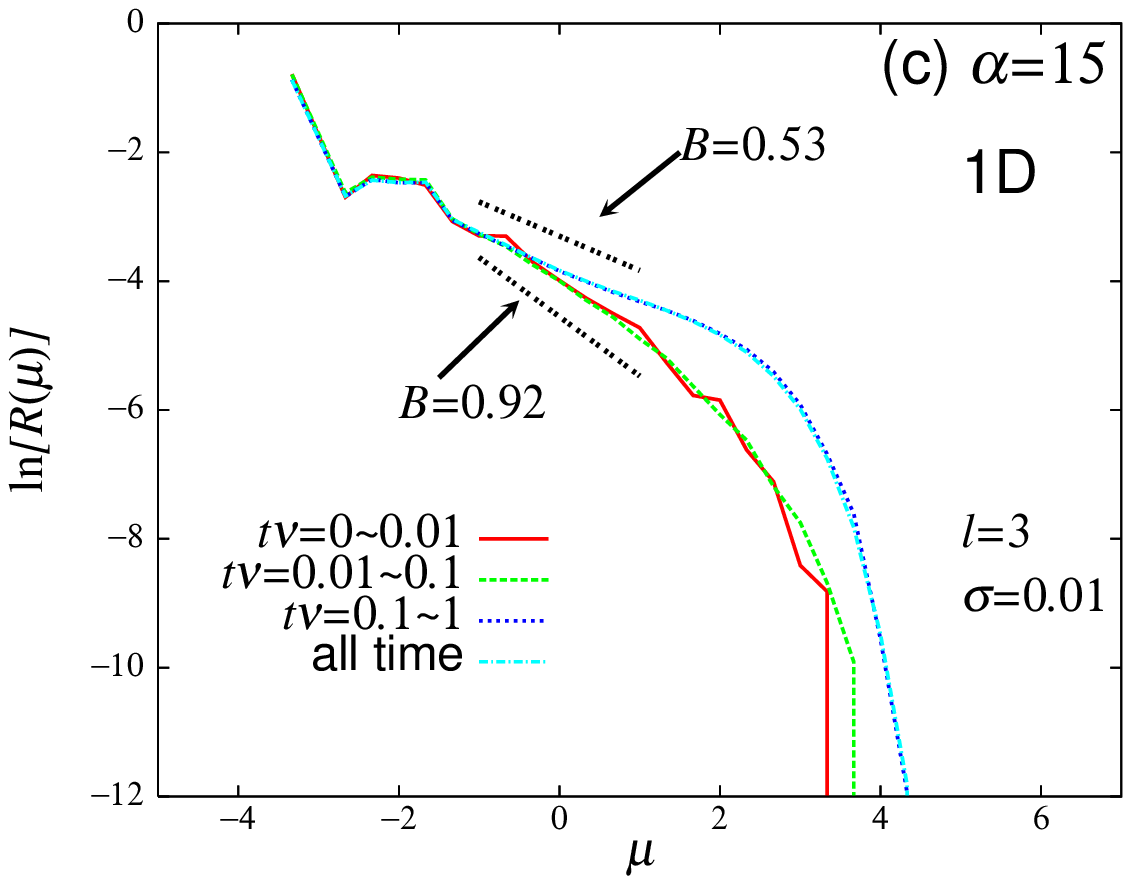}
\end{center}
\caption{
The local magnitude distribution of the 1D long-range BK model 
 for several time periods before the mainshock of $\mu >\mu _c=3$, for the cases of $\alpha=1$ (a), $\alpha=5$ (b), and $\alpha=15$ (c). 
Events whose epicenter lies within 30 blocks from
 the epicenter of the upcoming mainshock are counted.
The parameters $l$ and $\sigma$ are fixed  to  $l=3$ and 
$\sigma =0.01$. The system size is $N=800$.  In (a), the apparent $B$-value decreases before the mainshock, while, in (c), it increases before the mainshock.
}
\end{figure}

\end{document}